# Controversy and consensus: common ground and best practices for life cycle assessment of emerging technologies

Rachel Woods-Robinson[a], Amila Abeynayaka[b], Mik Carbajales-Dale[c], Hao Chen[c], Anthony Cheng[d], Gregory Cooney[e], Abby Kirchofer[f], Manish Kumar[g], Heather P. H. Liddell[h,i], Lisa Peterson[j,k], I. Daniel Posen[l], Sheikh Moni[m,n], Sylvia Sleep[o], Liz Wachs[p], Shiva Zargar[q], Joule Bergerson[r,s*]

[a] Clean Energy Institute, University of Washington, 3946 W Stevens Way NE, Seattle, WA 98195, USA
[b] Quantitative Sustainability Assessment, Department of Environmental and Resource Engineering, Technical University of Denmark (DTU), 2800 Kongens Lyngby, Denmark
[c] Environmental Engineering & Earth Sciences, Clemson University, Brackett Hall, Clemson, SC 29634, USA
[d] Department of Engineering and Public Policy, 5215 Wean Hall, Carnegie Mellon University, Pittsburgh, PA, 15232 USA, Carnegie Mellon University, Pittsburgh, PA 15213, USA
[e] Kleinman Center for Energy Policy, University of Pennsylvania, 220 S. 34th Street, Philadelphia, PA 19104
[f] Ramboll, 250 Montgomery Street, Suite 1200, San Francisco, CA 94104, USA
[g] Helmholtz Institute Ulm (HIU), Helmholtzstraße 11, 89081 Ulm, Karlsruhe Institute of Technology (KIT), Germany
[h] School of Mechanical Engineering, Purdue University, 585 Purdue Mall, West Lafayette, Indiana 47907 USA
[i] School of Sustainability Engineering & Environmental Engineering, Purdue University, 500 Central Drive, West Lafayette, Indiana 47907, USA
[j] Department of Civil, Architectural and Environmental Engineering, Drexel University, 3141 Chestnut Street, Philadelphia, PA 19104, USA
[k] Aftan Engineering, 839 Alleghenyville Rd, Mohnton, PA 19540, USA
[l] Department of Civil & Mineral Engineering, University of Toronto, 35 St. George Street, Toronto, Ontario, M5S 1A4 Canada
[m] National Energy Technology Laboratory (NETL), Pittsburgh, PA 15236, USA
[n] NETL Support Contractor, Pittsburgh, PA 15236, USA
[o] Department of Civil Engineering, Schulich School of Engineering, University of Calgary, 2500 University Drive NW, Calgary, Alberta, T2N 1N4, Canada
[p] National Renewable Energy Laboratory, 930 D St. SW, Suite 930, Washington, DC 20024, USA
[q] Department of Wood Science, Faculty of Forestry, The University of British Columbia, 2424 Main Mall, Vancouver, British Columbia, V6T 1Z4, Canada
[r] Department of Chemical and Petroleum Engineering, Schulich School of Engineering, University of Calgary, 2500 University Drive NW, Calgary, Alberta, T2N 1N4, Canada
[s] Department of Chemical and Biomolecular Engineering, University of Notre Dame, 250 Nieuwland Hall, Notre Dame, Indiana, 46556, USA

[*] Corresponding author: jbergers@nd.edu

## Abstract

Public and private interest in life cycle assessment (LCA) has grown as environmental disclosure norms tighten, driving demand for decision-relevant assessment early in technological development cycles. Early-stage LCA has the potential to guide design choices, steer innovation, and mitigate lock-in of adverse environmental impacts. However, many aspects of early-stage LCA practice remain unsettled. We convened experts in a series of Faraday Discussion-style workshops to address recurring debates across six key topics for emerging technologies: appropriate use of LCA, uncertainty, comparison with incumbents, standardization, scale-up, and stakeholder engagement. For each issue, we present a declarative resolution, summarize key arguments for and against it, identify points of consensus, and provide recommendations. Across topics, the research network converged on practical priorities including framing studies to the decision context; setting minimum reporting expectations for data and study quality; and explicitly stating limits of transferability for scenario-based uncertainty assessment or analytically scaled-up projections. Disagreements persisted on when to formalize standards and how extensively uncertainty can/should be treated for low-maturity technologies. Supplementing the workshop findings with examples and context from relevant literature, we synthesize outcomes into a set of shared challenges and research priorities to strengthen transparent, evidence-based, and context-informed approaches for early-stage LCA.

Keywords: Life cycle assessment (LCA); Life cycle thinking; Technology assessment; Design for environment (DfE); Eco-design; Emerging technology; Early-stage technology assessment

## Introduction

Life cycle assessment (LCA) can provide critical insights into the potential environmental impacts of emerging technologies across their full life cycle, from raw material extraction to manufacturing, use, and end-of-life. Emerging technologies (defined here as innovation concepts, materials, or products not yet commercialized) offer unique opportunities for proactive LCA, which could in turn inform decision-making. Design flexibility at early stages of development enables the integration of environmental considerations before decisions are locked in and real-world impacts occur. However, existing LCA frameworks and tools are not typically designed for such early-stage applications, posing a major challenge in adapting or developing appropriate methods for forward-looking assessments (Bergerson et al., 2020; Moni et al., 2019). While prospective approaches have been proposed, they are not yet standardized nor widely adopted, and important challenges remain.

The lack of clear guidelines for early-stage LCA has led to divergent approaches and opinions across the LCA community and ultimately has hindered integration of sustainability analysis at early stages. In response, researchers from the LCA of Emerging Technologies Research Network have been meeting regularly since 2017 to discuss and address challenges specific to emerging technologies. More than 100 members have participated in the research network over this period. In prior work (Bergerson et al. 2020), we outlined issues including data scarcity, uncertainty around industrial scale-up, lack of incumbent comparators, and unclear deployment or market pathways. The present study extends upon that foundation by responding to some of the key needs identified previously: to deepen dialogue, establish structured guidance where feasible, and build research networks to advance LCA practice and application at early development stages.

In attempting to systematically address the challenges to LCA posed by emerging technologies, we encountered deep disagreements about priorities and solutions. To describe this dynamic, we use the term "controversy"—not to suggest conflict or immaturity of the field, but rather to reflect ongoing debates and divergent perspectives. While these controversies are not always highlighted explicitly in the literature, our experience suggests that LCA practitioners (at least in North America) often hold strongly opposing views. For example, members of our research network have personally observed disagreements surfacing during recent workshops and sessions at the American Center for LCA (ACLCA) and International Symposium on Sustainable Systems and Technology (ISSST) conferences (Research Network Members, personal communications, 2023–2024). These discussions occasionally grew heated (e.g., around appropriate use of uncertainty techniques). Furthermore, conference polling conducted by our group revealed nearly even splits across all six of the debate resolutions that will be presented in this paper. Diverging views among this paper's authors and peer reviewers further reflect the field's active and unsettled nature.

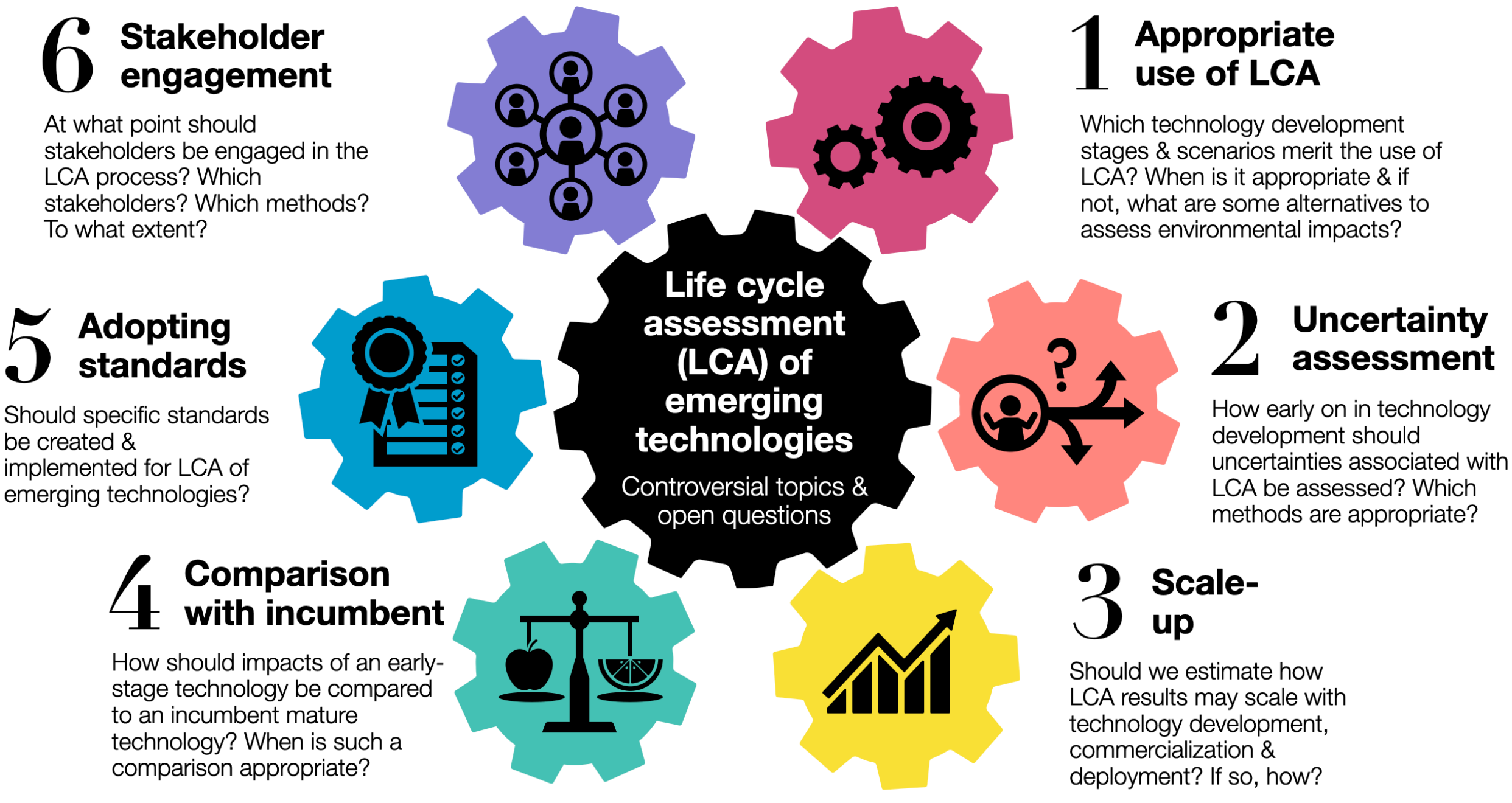

***Figure 1. Overview of controversial topics and open questions in the LCA of emerging technologies.*** *Note that the numbered items reflect the structure of the paper but do not imply a linear process flow. Each topic should be revisited iteratively throughout the stages of the LCA.*

Here, we focus on six key controversial topics central to LCA of emerging technologies, summarized in **Figure 1**: (1) appropriate use of LCA, (2) uncertainty, (3) scale-up, (4) incumbent comparison, (5) standardization, and (6) stakeholder engagement. Within each topic, we have identified a specific point of controversy. While not exhaustive, this list of controversies emerged as priority topics via both discussion and voting among >100 participants at a 2024 ISSST plenary session. In this paper, we do *not* perform a comprehensive methodological review of these topics (excellent examples already exist, e.g., Adrianto et al., 2021; Arvidsson et al., 2024; Buyle et al., 2019; Erakca et al., 2024; Florin & Romain, 2022; Marson et al., 2025; Moni et al., 2019; Nurdiawati et al., 2025; Thonemann et al., 2020), nor do we attempt to force consensus. Rather, our goals are to surface and synthesize debates to inform LCA practitioners, researchers, and stakeholders interacting with emerging technologies, and to make explicit the norms and methodological trade-offs often left implicit across individual studies.

Many of the methodological tensions underlying our six controversial topics, while long-standing within LCA in general, become especially pronounced for emerging technologies due to amplified uncertainties and limited data. Constructive debate is essential across LCA since it depends on models, assumptions, and normative choices regarding system boundaries, allocation, impact categories, and more. For example, prominent disagreements in the literature reveal differing foundational views on the validity and policy relevance of attributional and consequential LCA approaches (Plevin et al., 2014; Suh & Yang, 2014; Yang, 2016), and the appropriate use of hybrid vs. process-based approaches (Pomponi & Lenzen, 2018; Yang et al., 2017). Biogenic carbon accounting remains unsettled (Mayer et al., 2020), and plastics assessments are constrained by evolving development of characterization factors, especially for micro- and nano-plastics (Li et al., 2023, Corella-Puertas et al. 2023). Furthermore, across scientific fields "adversarial collaboration" (where researchers with opposing views engage as a team and jointly publish) offers an effective yet underused means to reduce bias and advance understanding (Ceci et al., 2024; Melloni 2023).

To navigate this debate-rich landscape, this paper draws on two complementary approaches. First, inspired by adversarial collaboration, we assembled a multi-author team to surface divergent perspectives on unsettled areas in prospective LCA, drawing on structured discussions and our recent open workshops. Second, we adopt a Faraday Discussions-style format (originally developed for physical chemistry) to structure each topical section around a brief literature overview, a proposed "controversial resolution," two opposing viewpoints, areas of agreement, and key takeaways. Each topic was examined by subgroups within the network to distill the core controversy, summarize major arguments for and against resolutions, and support opposing positions with literature (much of it recent, highlighting the timeliness of these issues). Throughout, uncited statements reflect insights raised in workshops, while cited material represents documented positions in the literature.

To contextualize each topic, every section includes a summary figure that follows a consistent format, with the goal to foster constructive dialogue. Each figure frames topics along a qualitative continuum of technology development, from concept to commercial scale, inspired by Technology Readiness Levels (TRLs). Throughout, we use "TRL" as shorthand for broad qualitative stages of technological maturity rather than precise numerical values (unless quoting literature). Figures are intended as *qualitative* interpretive guides: elements provide relative (not absolute) relationships that should be explored further, while axes represent qualitative (not quantitative) and sometimes composite dimensions commonly used in heuristic frameworks (Osgood et al., 1957). Figures are *not* intended as consensus recommendations; while we did not agree on definitive "right answers" to open controversies, we reached consensus on key points practitioners should consider when conducting early LCA.

Lastly, we highlight core challenges, practitioner questions, and strategic recommendations at the end of each topical section. As an orientation resource to guide readers, these key takeaways are also compiled in aggregate in **Table 1**, which we intend as a synthesis and standalone reference of open controversies, areas of consensus, considerations for LCA practitioners, and next steps for the field. The paper concludes with a reflection on the implications of these debates, offering guidance to LCA practitioners and stakeholders to raise awareness of common challenges in assessing emerging technologies and to support informed dialogue on how to navigate them.

***Table 1****. Summary of topic questions, key challenges and takeaways, recommendations for practitioners, and next steps in the field. The term technology readiness level (TRL) is used qualitatively to refer to broad developmental stages rather than specific TRL numbers (1–9). While several of the statements apply to LCA in general, we highlight them as specifically pertinent to early-stage LCA given the unique characteristics of emerging technologies (ETs). These special characteristics include high uncertainty and unclear pathways to commercialization.*

| **Topic**<br>W/in ET LCA | **Open "controversies"**<br>Questions to resolve | **Key "consensus" points**<br>Challenges & takeaways from discussions | **Recommendations for practitioners**<br>Considerations for conducting early-LCA | **Next steps in the field**<br>Research, industry & policy priorities |
|---|---|---|---|---|
| **Appropriate use of LCA** | • How early is "too early" to use LCA?<br>• Should screening-level LCAs still be called "LCA"?<br>• How should LCA robustness be assessed & communicated?<br>• How do we prevent misuse of screening or early-stage LCA results? | • **Life cycle thinking ≠ robust LCA.** This distinction is critical when sharing/applying results.<br>• **TRL data vs. design tradeoffs.** Data availability increases with TRL, but design flexibility (to mitigate impacts) diminishes.<br>• **Delaying LCA has risks** such as lock-in, missed hotspots & suboptimal R&D priorities.<br>• **Early LCA can also have risks**. Failure to disclose limitations can mislead decision-makers & undermine trust. | • **Clarify intended use.** Explicitly state purpose, boundaries, limits & decision context to avoid misinterpretation.<br>• **Assess LCA robustness.** Clearly convey whether/how study supports intended use given goal & scope.<br>• **Align method depth to maturity.** Choose approaches appropriate to TRL & data quality to avoid overreach.<br>• **Use screening LCAs mindfully.** Simplified methods (e.g., spotlight risk analysis) may help at low TRL; but clearly state they are not robust LCAs. | • **Develop consistent terminology** to classify early-stage LCA rigor; clearly communicate strength & limits of conclusions.<br>• **Create tiered frameworks** for LCA aligned with TRL, data quality & decision context.<br>• **Design early-stage tools** which can evolve into robust LCAs as data improves.<br>• **Set robustness guidelines** for when early-stage LCA rigor is sufficient to support specific decisions. |
| **Uncertainty assessment** | • At which TRL should uncertainties begin to be assessed?<br>• Which methods are appropriate at each TRL?<br>• How should results be communicated to avoid misinterpretations? | • **Uncertainty is intrinsic to LCA** & even more pronounced for ETs due to data gaps, etc.<br>• **Effective communication** of uncertainty in at least some capacity is crucial (e.g., explicit reporting, contextualizing, scenario development, visual aids).<br>• **Dynamic evolution**: LCA uncertainty methods need to adapt to the evolving nature of ETs. | • **Report methods transparently**. Adopt clear & comprehensive practices for communicating uncertainty.<br>• **Use probabilistic methods cautiously.** Ensure appropriate data & context for advanced modeling.<br>• **Consider bounding estimates**, performance thresholds or breakeven analyses at lowest TRLs.<br>• **Focus on sensitivity.** Prioritize uncertainty drivers that lead to hotspots rather than absolute results. | • **Advancement of methods**. Continue to develop methods to characterize, quantify & propagate uncertainty in data poor contexts.<br>• **Contextualize method selection**. Develop consensus on how to best align uncertainty approaches with TRL, data availability & study objectives. |

| Topic<br>W/in ET LCA | Open "controversies"<br>Questions to resolve | Key "consensus" points<br>Challenges & takeaways from discussions | Recommendations for practitioners<br>Considerations for conducting early-LCA | Next steps in the field<br>Research, industry & policy priorities |
|---|---|---|---|---|
| **Scale-up** | • Should methods be applied to project future performance of ETs?<br>• If so, which methods & learning types should be considered?<br>• At what TRL or data maturity is scale-up modeling meaningful? | • **Rigor vs. risk trade-offs.** Inappropriately scaling low-TRL systems may yield new uncertainties & false confidence.<br>• **Nonlinear scale-up:** Material, process, efficiency & infrastructure changes are rarely linear. | • **Apply context-aware scale-up methods** based on TRL, development trajectory & data availability; align LCA with realistic process design & learning effects.<br>• **Explain assumptions** & provide evidence that the scale-up framework is appropriate for the technology assessed. | • **Review current scaling frameworks** & modify to address gaps.<br>• **Link to process modeling:** Work with engineers to build dynamic scale-up models.<br>• **Create open-access scale-up databases** of processes across technologies, markets & TRLs.<br>• **Develop user-friendly tools** to help practitioners apply scale-up. |
| **Scale-up & comparison** (applies to both) | | • **Results depend on framing,** e.g., comparator choice, incumbent vs. ET TRL & scale-up approach.<br>• **No ISO guidance** or consensus on how to fairly project & compare tech at varying TRLs. | • **Evaluate comparison validity**. If LCA is comparative, consider TRL of each technology system & whether a scale-up method is needed for the ET to make an appropriate comparison. | |
| **Comparison with incumbent** | • Should comparisons be made at low TRL?<br>• If so, how should an appropriate comparator be determined?<br>• How should scaling dynamics, uncertainty, system boundaries & bias be accounted for? | • **Comparison often implied**. LCAs of ETs typically have an incumbent comparison in mind, even if not explicitly comparative.<br>• **Benchmarking can add value** & aid progress tracking but requires careful contextualization. | • **Clarify comparison intent**. E.g., benchmarking, informing R&D, identifying trade-offs, etc.?<br>• **Audit implications & potential biases of comparator choices**. E.g., single vs. mixed incumbents, data sources, modeling, alternate functions, future scenarios. | • **Build consensus on comparator selection,** appropriate methods (e.g., marginal, average, best-in-class) & characterization approaches.<br>• **Encourage scenario-based approaches** to reflect uncertainty in both ET & incumbent.<br>• **Clarify handling of forecast uncertainty.** Set norms on how to address uncertainties in ET future performance & incumbent evolution |
| **Standard-ization** | • Could/should additional standards play a role in assessing ETs?<br>• If so, what should they standardize? Methods, assumptions, reporting?<br>• Should they be guidelines or standards? Sector-specific? | • **Standards offer value** for ET LCA. E.g., reduce *ad hoc* nature of current method decisions.<br>• **Standards also have pitfalls.** E.g., overconfidence in accuracy & comparability; procedural burden.<br>• **Guidelines set in some sectors** but no general standards or requirements.<br>• **Diverse tech & contexts** make setting universal ET standards challenging. | • **Start with existing ISO guidance** from 14040 series when feasible.<br>• **Apply sector-specific guidance** if emerging standards are available & relevant.<br>• **Explore impact of different modeling choices** when standards insufficiently constrain methods. | • **Explore community consensus** on what should be standardized, where flexible guidance is most appropriate & where sector-specific guidance is needed.<br>• **Continue development** of standard sets of narratively consistent scenarios for future background systems. |

| Topic<br>W/in ET LCA | Open "controversies"<br>Questions to resolve | Key "consensus" points<br>Challenges & takeaways from discussions | Recommendations for practitioners<br>Considerations for conducting early-LCA | Next steps in the field<br>Research, industry & policy priorities |
|---|---|---|---|---|
| **Stakeholder engagement** | • Should stakeholders be included in LCA of ETs?<br>• If so, which stakeholders?<br>• When should engagement occur?<br>• What is the appropriate degree of engagement? | • **Explicit "engagement" guidance is lacking** in ISO 14040 series about when/how to engage stakeholders in LCA.<br>• **Diverse perspectives can provide value** in LCA design & also throughout the study.<br>• **Subject matter expert (SME) involvement enhances rigor**. SMEs can assist scoping, feasibility checks, blind spot identification, etc.<br>• **Community input improves relevance** & equity by informing assumptions, priorities, local concerns, etc. | • **Explore potential stakeholder roles**, e.g., when/how stakeholders could/should contribute to LCA of ETs.<br>• **Engage early & intentionally**. Ensure consultation (if relevant/planned) occurs during the scoping phase & prior to associated lock-in.<br>• **Avoid superficial inclusion.** Ensure that engagement is meaningful, transparent, & adequately resourced to prevent tokenism or overburdening stakeholders.<br>• **Examine potential analyst biases** & blind spots; avoid value judgments masquerading as expertise. | • **Develop engagement guidance.** Outline pros/cons & appropriate levels of involvement (informing to empowering) aligned with LCA type, TRL & stakeholder group.<br>• **Systematically analyze case studies** that have engaged stakeholders to assess outcomes, benefits, challenges & best practices.<br>• **Promote community dialogue** on stakeholder roles in LCA to build consensus & capacity.<br>• **Minimize burden** on analysts & communities through carefully targeted approaches. |
| **Overarching themes** | • How can the community ensure coherence & transparency across early-stage LCA practices? | • **Screening LCAs can/do guide R&D** by identifying hotspots & gaps, but limits must be clarified.<br>• **Fragmented practices**. Inconsistent methods limit comparability & trust in early-stage LCA.<br>• **Key ET LCA trade-offs**: flexibility vs. rigor, comparability vs. contextual relevance. | • **Contextualize conclusions** (e.g., "X true only if Y") with assumptions & scope; avoid claims beyond what data can meaningfully support.<br>• **Ensure transparent reporting.** Even if ET-specific standards or guidance is lacking, clearly state assumptions & limitations in interpretation. | • **Set consistent methods guidance** on context-dependent considerations (scaling, etc.)<br>• **Establish communication norms** & a minimum baseline for reporting results—appropriate use, uncertainty, scale-up, etc.—using clear language, context & visuals. |

## Topic 1: Appropriate use of LCA

The first controversial topic centers on a core dilemma: whether, when, and how it is appropriate to apply LCA to emerging technologies. On one hand, there is broad agreement that LCA can most effectively advance sustainability goals when implemented early in technology development, before environmental impacts become locked in by design and deployment choices (Tischner & Charter, 2001; Cucurachi et al., 2018; Majeau-Bettez, 2021; Wender et al., 2014). On the other hand, meaningful LCA requires a definable system and adequate process data. As Grimaldi (2020) cautions, early-stage LCAs must be treated carefully. When uncertainty about feedstocks, process yields, or deployment scenarios is high, results may mislead, overstate benefits (and therefore risk misdirecting decision-makers to entrench unsustainable pathways), or cause premature abandonment of promising options (Moni et al., 2019).

This creates a central tension: apply LCA too early and risk misleading results, or apply it too late and risk losing opportunities to guide sustainable design. Our discussions reflected divergent views on how to address this, reinforcing the need for approaches that align analytical depth with technology maturity, maintain methodological rigor, and clearly communicate scope and limitations. **Figure 2** illustrates how LCA practices and challenges shift across TRLs. At later stages, more complete inventories and context enable comprehensive assessments. At very early stages, where full LCA is often infeasible (e.g., due to data-poor contexts), simplified “life cycle thinking” approaches (Todd & Curran, 1999)—qualitative or semi-quantitative consideration of impacts across life cycle stages—may be more suitable. These alternatives, such as semi-quantitative mapping (Hung et al., 2018; Majeau-Bettez, 2021), risk matrices (Horgan et al., 2022), and screening tools like “knockout” or “stoplight” diagrams (Putsche et al., 2023; Todd & Curran, 1999), can identify potential environmental hotspots in concept- or lab-scale designs, set targets of performance (e.g., general criteria that are life cycle informed). Though limited in scope and infrequently reported, such screening tools can support early decisions and help scope future LCAs even when they might not be suited for rigorous uncertainty or scale-up analysis.

The mid-stages of technology development offer an opportunity for what we term “early-stage LCA frameworks” that balance limited data with growing technological definition (see Figure 2). These include LCA methods that integrate process simulations, learning curves, expert judgment, and/or scenario modeling to inform projections (Marson et al., 2025). Applied consistently, these frameworks could support decision-making such as technology risk ranking, helping funders, regulators, and developers prioritize options. Although numerous variants exist with varying definitions, including prospective (Thonemann et al., 2020; Arvidsson et al., 2024; Nurdiawati et al., 2025), anticipatory (Wender et al., 2014), and ex-ante LCA, including scenario-based frameworks (Cucurachi et al., 2018; Buyle et al., 2019), no standard terminology or best practices have emerged (Grimaldi, 2020; Marson et al., 2025). Buyle et al. (2019), for instance, reviewed over 50 early-stage “ex-ante” LCAs with varying assumptions, demonstrating the field’s fragmentation. This underscores the need to build consensus, align analytical rigor with data availability, and clearly convey assumptions, limitations, and intended uses.

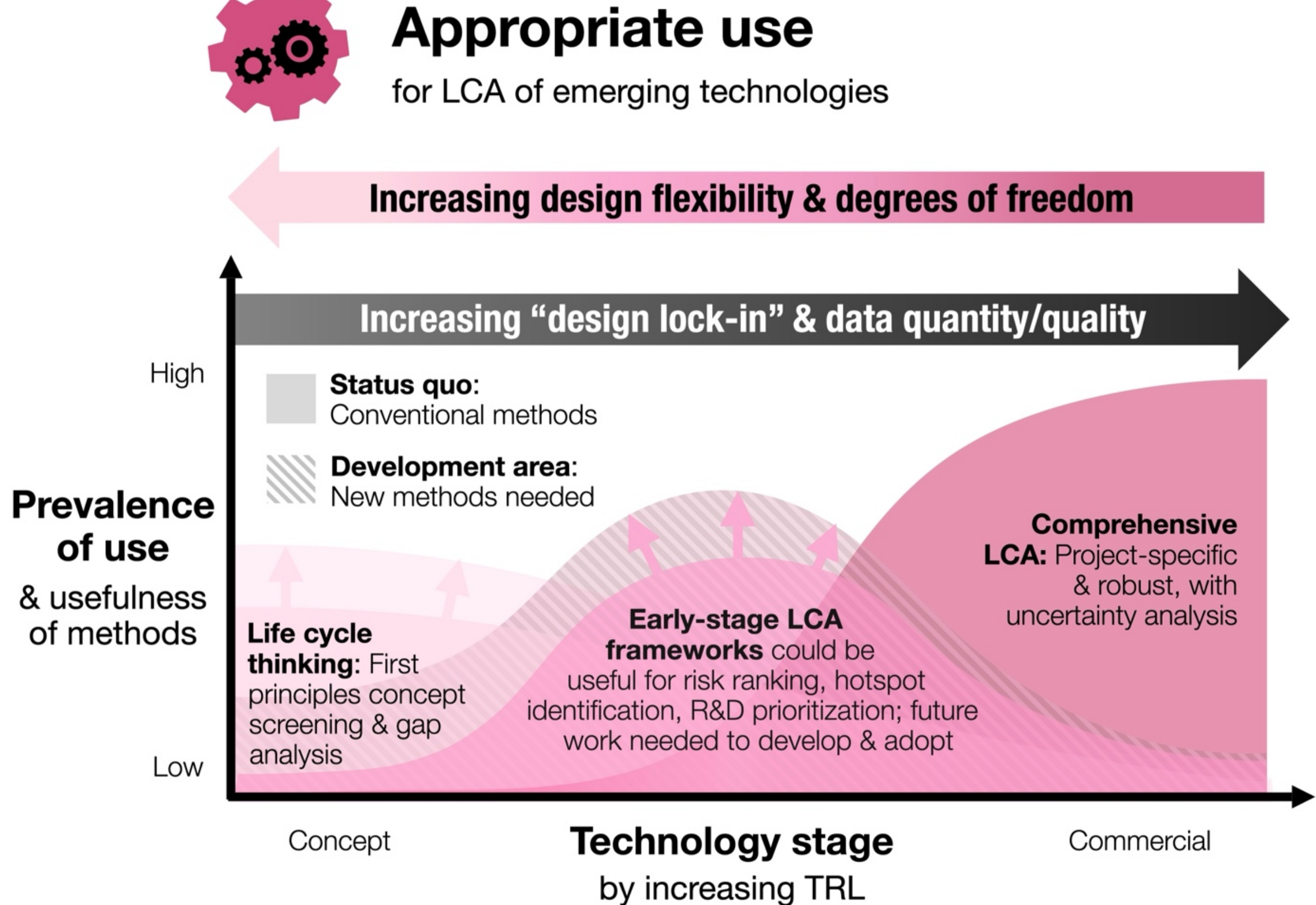

***Figure 2. Appropriate uses of LCA across technology development stages.*** *This schematic illustrates the evolving relevance and application of life cycle–based analysis approaches across increasing technology readiness levels (TRLs). The y-axis reflects both the current or potential use and the perceived appropriateness of life cycle–based approaches at across TRLs, recognizing that some approaches, though not yet widely used, offer value and could benefit from further development and adoption. At low TRLs, where design flexibility is greatest, LCA is rarely applied (indicated by smaller curve heights). Conversely, at high TRLs, LCA is more commonly performed (larger curve heights), though the potential for design "lock-in" also increases. The hashed shading denotes potential opportunities for methodological development and broader adoption, particularly in mid-development stages, where more system data becomes available and deployment contexts become clearer. Arrows in the figure indicate opportunities for more integration of life cycle thinking methods early on (closer to concept phase), provided they are clearly distinguished from comprehensive and robust LCAs. Note that Figures 2–7 are illustrative: they represent feedback from workshops and author discussions, rather than prescriptive guidance or definitive consensus, while curve shapes and heights are qualitative, guided by the experience and expertise of co-authors.*

## Controversial resolution: LCA is always an appropriate and helpful tool, even at early development stages.

## Support for the resolution

There is growing consensus that some form of life cycle thinking adds value across all development stages. Even under high uncertainty, LCA can guide gap analysis by identifying missing data, key sensitivities, and environmental hotspots. When interpreted carefully, these insights can shape R&D priorities and support more sustainable decisions. While comprehensive LCAs may not be feasible early on, these core concepts remain valuable if uncertainties are transparently communicated.

Specifically, early-stage assessments can identify burden shifting and environmental trade-offs by taking a systems view, even with limited data. Methods like scenario analysis, learning curves, and proxy data from similar technologies help fill data gaps (Gavankar et al., 2015; Piccinno et al., 2016; Weyand et al., 2023). New risk-based tools (e.g. the Emerging Materials Risk Analysis in Horgan et al. (2022)) and early-stage screening metrics (National Research Council, 2009) can also provide decision-relevant insights. For instance, Putsche et al. (2023) applied stoplight diagrams to TRL 1 battery materials, identifying red flags such as hazardous reagents and critical precursors before lab work began. Although energy and process routes were unknown, emerging data-driven methods were used to estimate these parameters in support of early modeling to support early-stage design choices. Such efforts are not comprehensive LCAs, but they lay the groundwork for them. They also make it easier to iterate as better data become available.

In this way, life cycle thinking should contribute across TRLs by:

1. **Preventing design lock-in** before key decisions are fixed.
2. **Highlighting trade-offs** or hotspots that might otherwise go unnoticed.
3. **Guiding R&D** toward the most environmentally promising pathways.

## Opposition to the resolution

While early LCA is increasingly requested, applying it without robust data or defensible assumptions can lead to misleading outcomes. LCAs included to satisfy funding expectations may fall short of minimum standards (Brandão et al., 2024; Rebitzer & Schäfer, 2009). If limitations aren't clearly disclosed, such results can distort comparisons, mislead decision-makers, or undermine credibility. For example, screening LCA results risk being misinterpreted as definitive by non-experts unfamiliar with their limitations.

Potential risks include:

1. **Misleading decisions**: Not considering uncertainty in a systematic way can lead to a false sense of *certainty* in LCA impact estimates. Over- or underestimation of impacts can lead to entrenching unsustainable pathways or abandoning pathways prematurely because the full potential has not been captured in the analysis. (See Topic 2.)
2. **Faulty comparisons**: LCA findings are highly sensitive to assumptions. Comparing emerging technologies to incumbents or to one another can yield misleading results when datasets differ in quality or completeness. While target-setting may be possible, direct comparisons may not be appropriate.

## Common ground

We agree that life cycle thinking is valuable at all stages but is not equivalent to a robust and comprehensive LCA. As **Table 1** notes, distinguishing between screening-level assessments and comprehensive studies is critical. Following ISO 14040 guidance on consistency, completeness, transparency, and logical conclusions aligned with the study's goal and scope are important starting points, particularly for emerging technologies where assumptions and uncertainties dominate. For early-stage technologies, articulating the study's goal and scope—including defining a plausible function and functional unit—is more complex and critical. Yet ISO does not advise how to determine whether the study's findings are sufficient to support that decision. Guidance on how to judge whether an LCA meets that threshold and how to present evidence accordingly would improve result interpretation and application.

A useful strategy may involve classifying LCAs into tiers of rigor, similar to how techno-economic analyses (TEAs) are scaled by technology maturity, complexity, and data quality (Bredehoeft et al., 2020). This would clarify expectations and align methods with a study's purpose and the maturity of the technology assessed. In sum, careful communication, methodological alignment with technology maturity, and transparency in scope and limitations are key to responsibly applying LCA early in development.

## Topic 2: Uncertainty

Uncertainty in LCA is pervasive, even for mature technologies (Heijungs, 2024; Igos et al., 2019), and becomes more acute when applied to emerging technologies. Early-stage technologies lack historical data, their potential future performance (foreground system) is unknown, and depend on speculative background systems (e.g., energy supply). These background systems, modeled using secondary databases, often fail to capture major societal, infrastructure, or energy transitions. In addition to standard sources of LCA uncertainty (e.g., temporal or geographic variability), early-stage technologies face high scenario uncertainty (unknown future conditions), parameter uncertainty (technological options), and model uncertainty (subjective modeling choices) (Blanco et al., 2020). **Figure 3** illustrates how different uncertainties and methods evolve across TRLs, showing that while uncertainty is greatest early on, the ability to apply well characterized and sophisticated methods increases later, revealing a critical mismatch.

While many uncertainty assessment methods exist (e.g., Heijungs et al., 2019; Saltelli et al., 2007; Weidema & Wesnæs, 1996), few are tailored to the complexities of early-stage technologies. Recent research has begun to address this gap. For instance, Blanco et al. (2020) present a probabilistic method to emerging photovoltaic systems and Jouannais et al. (2024) present a probabilistic evaluation in ex-ante LCA generally. Grimaldi (2020) shows how iterative LCAs aligned with process development can reduce LCI uncertainty and benchmark against incumbents, though scaling challenges at low TRLs can still render projections unreliable. Douziech et al. (2023) integrate parameterized life cycle inventories (LCIs) of foreground and background systems using open-source tools to address background system evolution and enable robust comparisons, while others explore economic and policy influences, such as early-stage adoption prices (Miller et al., 2020) and IAM-driven LCI databases such as Premise (Sacchi et al., 2022).

Still, as Figure 3 suggests, there is no consensus on best practices for uncertainty in early-stage LCA. A key dilemma is that uncertainty is highest when tools to assess it are least reliable. This leads to divergent views. Some advocate for more complex methods (e.g., Monte Carlo simulation) to quantify the high levels of uncertainty in early-stage technologies (e.g., Blanco et al., 2024; Jouannais, 2024), while others argue that uncertainty at low TRLs is too deep for such tools to be meaningful given that uncertainty assumptions themselves are likely to be flawed (e.g., Heijungs, 2024).

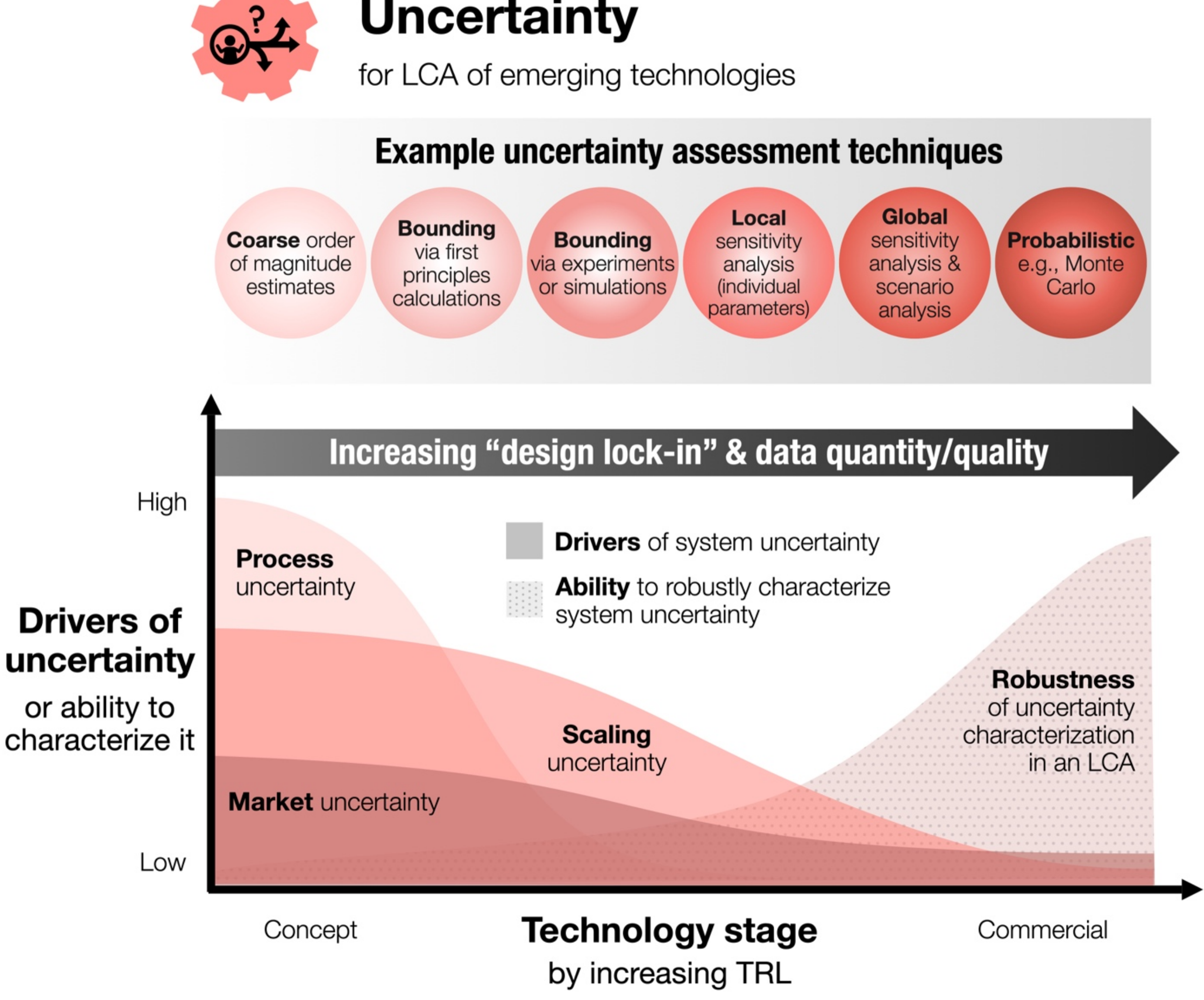

***Figure 3. Uncertainty considerations and assessment methods for LCA of emerging technologies across technology readiness levels (TRLs).*** *The top panel presents examples of uncertainty techniques, loosely arranged by their methodological sophistication (we intentionally avoid ordering techniques by TRL, acknowledging differing views among authors on their appropriate sequencing or application range). Bounding refers to determining basic ranges of LCA estimates (upper and lower limits) with minimal assumptions. The bottom panel showcases either (1) non-exhaustive examples of prominent uncertainty types (pink shades), or (2) the ability to apply complete or sophisticated uncertainty assessment (dotted fill). Process uncertainty refers to the overall process design and is often a key uncertainty driver at the earliest stages; scaling uncertainty refers to challenges estimating commercial-scale performance from laboratory or pilot scale data and is frequently a critical uncertainty driver at mid-development stages; market uncertainty refers to uncertainty in ultimate product function, the specific market(s) into which the technology will be deployed, or how users or other related technologies and systems will ultimately interact with the product – its prominence varies heavily across product-types. At early stages, uncertainty is high across the board, and assessments are often limited to basic techniques. Toward commercialization, uncertainties can be characterized and quantified more comprehensively. At the same time, the potential for both design and technology "lock-in" increases as the degrees of freedom reduce. Note that Figures 2–7 are illustrative: they represent feedback from workshops and author discussions, rather than prescriptive guidance or definitive consensus, while curve shapes and heights are qualitative, guided by the experience and expertise of co-authors.*

## Controversial resolution: LCA of emerging technologies requires sophisticated uncertainty assessment.

## Support for resolution

Uncertainty is present at all stages of technology development, whether we acknowledge it or not (Gavankar et al., 2015). Therefore, conducting a structured uncertainty assessment of LCA results (e.g., generating a probability distribution), even if based on somewhat arbitrary assumptions, helps communicate inherent variability/uncertainty. This avoids presenting results as point estimates (single values without uncertainty bounds) that can misleadingly suggest unjustifiable certainty. Simply put: if an LCA practitioner does not believe that a point estimate can stand alone as a single "best guess" or central case based on experience and intuition, then it shouldn't be presented as such. Uncertainty assessment using a formal method such as Monte Carlo simulation allows the LCA practitioner to quantify (in a systematic way) the degree and nature of uncertainty and implications for the ultimate performance of the technology.

Communication of results using confidence intervals can be a useful way to succinctly capture and quantify the magnitude of uncertainty (Igos et al., 2019). It can also be a key tool for decision making such as when evaluating the likelihood of meeting certain thresholds of performance. For example, even at the very lowest TRL levels, first principles calculations combined with order of magnitude estimates can help identify theoretical upper and lower limits to physical and chemical parameters that can inform performance targets and life cycle inventories. While simple ranges or best-worst bounding scenarios can capture optimistic and pessimistic outcomes, probabilistic methods enable tighter and more realistic bounds on the results. Such methods can be more descriptive than artificially setting all parameters to their best or worst values and examining edge cases through parametric sensitivity approaches as they often involve arbitrary combinations of high/low across multiple inputs.

## Opposition to resolution

As highlighted in Figure 3, the same data gaps that drive uncertainty in early-stage LCA also hinder robust characterization of the nature and magnitude of that uncertainty across a system's life cycle. Thus, in aiming to prevent overconfidence in the point estimate, a detailed uncertainty assessment can unintentionally mislead stakeholders or decision-makers due to over-confidence implied by the final probability distribution (Heijungs, 2020). At a minimum, it should be explicitly acknowledged and communicated that such models suggest likely tendencies—not definitive outcomes—even when results are presented probabilistically.

For early-stage technologies, the key question is often not how well we expect the technology to perform, but rather does it have the *potential* to outperform the incumbent; or what choices in the technology's development are likely to drive its environmental performance. These are questions which may be fundamentally better served with simple bounding exercises or scenario analyses to highlight the conditions under which it does or does not perform well. For example, Saltelli et al. (2013, 2020) propose sensitivity auditing for environmental modeling: essentially a practitioner checklist that helps practitioners to highlight assumptions, check for inappropriate use of models that can come from over-interpreting results, and confirm that the correct questions are being asked.

Probabilistic models are best suited to capture parametric uncertainties (i.e., known unknowns), which may pale in comparison to the uncertainty in model structure (e.g., the function the technology serves, future projections, scale-up from laboratory data) that frequently dominate emerging technologies. When it comes to emerging technologies, we are often navigating in the realm of "unknown unknowns," where the future is unpredictable and uncertainties are not just unknown but also unquantifiable. This additional layer of uncertainty introduces profound complexities in quantifying future uncertainties, making it challenging to apply traditional LCA methods effectively. Thus, qualified point estimates (informed by bounding exercises or scenario analysis) may instead be more helpful (Villares et al., 2016).

### Common ground

Uncertainty is inherent in all LCAs, even for mature technologies, and requires careful treatment, particularly when applied to emerging technologies where deep uncertainties exist about design and potential impacts. When presenting LCA results, regardless of the approach to treat uncertainty (i.e., use of Monte Carlo simulation versus order of magnitude or bounding analyses), discussion should focus on factors driving environmental impacts or hot-spot type analyses rather than a focus on the absolute or precise quantitative results. Sophisticated probabilistic methods should be applied with caution. Furthermore, clear and effective communication of the limitations of the study given the early stage of the technology can prevent the study audience from misinterpreting either point estimate results or probability distributions as an absolute prediction of the future environmental performance of the technology. Effective communication of uncertainty should include consideration of explicit reporting, contextualization, scenario crafting, the use of straightforward language, and visual aids in the interpretation stage of the study. Given our group's disagreements on appropriate treatment of uncertainty, we call for clearer guidance on how to model and communicate uncertainty for emerging technologies, tailored to factors such as technology stage, uncertainty type, available resources, and the study's goal and scope. What we do agree on is that new methods to characterize, quantify and propagate uncertainty in LCA, especially with regards to poor data conditions typical of emerging technologies, could greatly improve the use and interpretation of LCA results.

## Topic 3: Scale-up

Understanding how a technology evolves through development stages and how to incorporate these changes into LCA remains unresolved. Tsoy et al. (2020) define "upscaling methods" as procedures that project how a technology currently at a lower TRL might look and function at a higher TRL (not to be confused with scaling the use of LCA itself). These challenges arise both in background systems (e.g., electricity mix) and foreground systems (e.g., technology development), often reflecting general issues in technology forecasting rather than problems unique to LCA. While ISO 14040 series standards do not explicitly recommend scale-up frameworks, ISO 14044 (B.3.4 Consistency check) highlights inconsistencies such as comparing LCAs based on experimental processes (Option A) with mature large-scale technologies (Option B). This inconsistency has motivated calls to better characterize scaling effects and to develop data-driven, rational methods to scale LCA results.

Interest in scale-up has grown considerably over the past decade, with multiple studies outlining relevant scaling mechanisms, modeling approaches, and associated challenges. Effective modeling begins with identifying current technology stage and how systems may evolve at scale. Although no universally accepted scale-up *procedures* exist, the literature offers some consensus on defining scaling *mechanisms or archetypes* across development stages (Riondet et al. 2023, 2024). Van Der Hulst (2020) categorized mechanisms into technology development (process changes, size scaling, process synergies) and industrial development (industrial learning and external factors). Weyand et al. (2023) expanded this into three learning types—process learning (technological learning, size scaling, industrial learning), material learning (new systems, substitutions, optimization)—depicted across corresponding TRL ranges in **Figure 4** (upper panel).

A variety of scale-up models have been proposed and applied to address these mechanisms and stages. Tsoy et al. (2020) reviewed early-stage ex-ante LCA upscaling methods, categorizing them as process simulations, manual calculations, molecular structure models, and proxies. Erakca et al. (2024) further distinguish extrapolation from empirical scaling and introduce modular influence estimation. Reviewing 78 studies, Erakca et al. (2024) identified 14 systematic scaling methods and created a tool to help modelers select appropriate methods based on context. Early work focused on specific sectors, such as chemical processes (Shibasaki et al., 2007; Piccinno et al., 2016, 2018). Weyand et al. (2023) introduced UpFunMatLCA, a structured framework for projecting future development and best-case upscaling scenarios for emerging functional materials ("lab to fab"). Figure 4 illustrates examples of technology changes through development stages and typical methods to quantify impacts, prevalence, and emerging innovations.

Despite extensive research, no consensus exists on when or how to apply scale-up methods. Early-stage technologies at lab, bench, or pilot scale are often inefficient in terms of material and energy use, and unrepresentative of future industrial-scale systems (Cucurachi & Blanco, 2022; Moni et al., 2019). Mass balance and life cycle inventory differences complicate fair comparisons with mature technologies (see Comparison to Incumbent section). Practical limitations—limited sector-specific guidance, scarce data on scale-up factors (especially for novel unit processes), proprietary data, uncertainties in learning curves, and a gap between LCA and process engineering training—hinder consistent application (Majeau-Bettez, 2021; Piccinno et al., 2016; Riondet et al., 2024). The field grapples with whether the complexity and uncertainty of scale-up frameworks produce sufficiently improved estimates to justify their use.

### Controversial resolution: Scale-up methods should be applied to early-stage LCAs to forecast performance at higher TRLs.

### Support for resolution

Scale-up methods are crucial to maintain the relevance and representativeness of LCAs for emerging technologies. As technologies progress from lab to industrial scale, LCA results change due to variations in materials, unit operations, process equipment, material and energy efficiencies (which stabilize at higher maturity), and resource utilization strategies (e.g., waste heat recovery, recycling). These shifts become more pronounced in comparative LCAs, especially when early-stage technologies are benchmarked against mature commercial systems (some operational for decades; see Comparison to Incumbent). Lab-scale setups differ fundamentally from industrial systems; thus, scaling inventory data is essential for plausible comparison across TRLs.

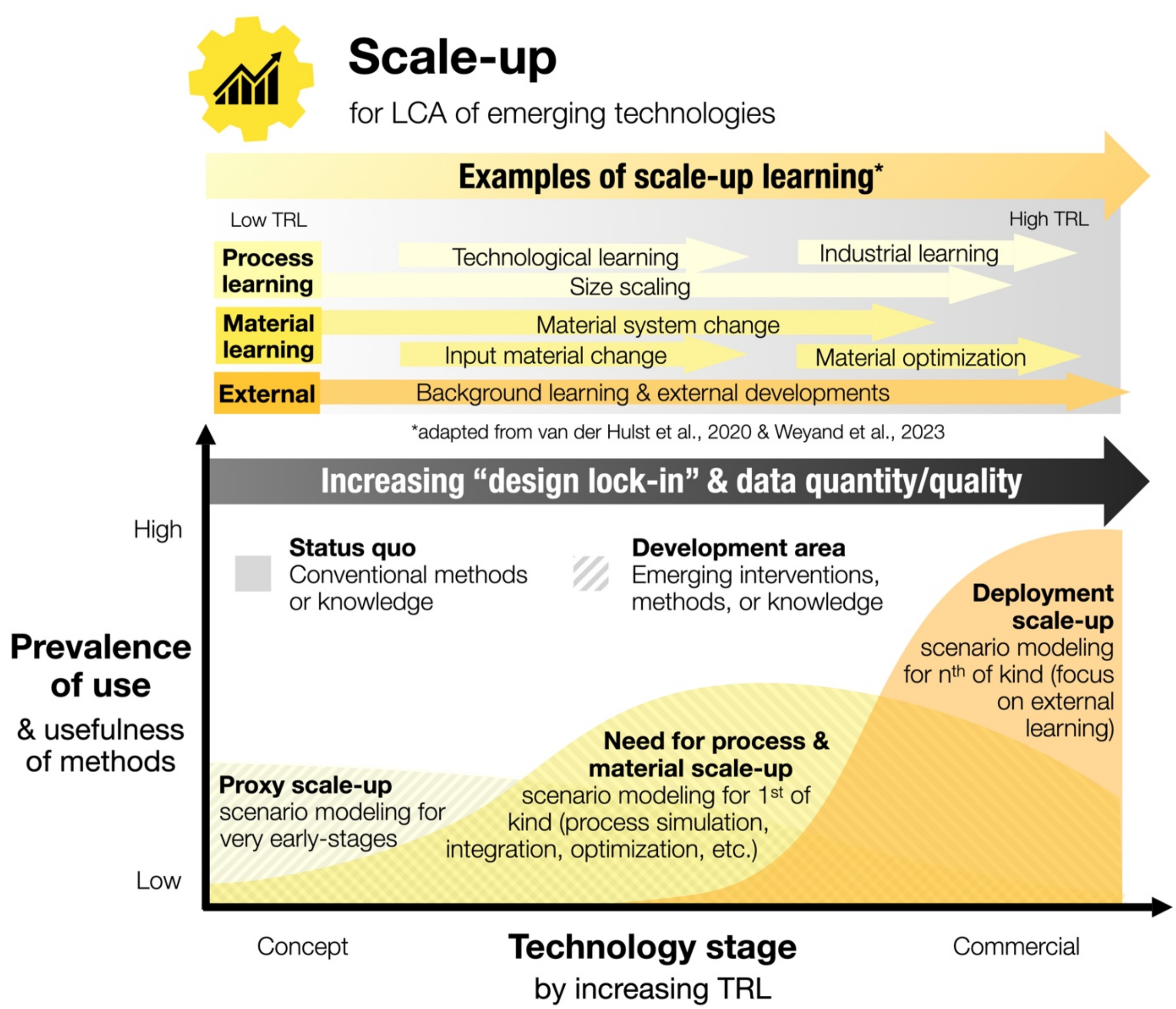

***Figure 4. Scale-up methods and considerations of systems in LCA across technology stage****. In LCA for emerging technologies, it is important to assess and project how a technology might change as a function of development stage, such as technology readiness level (TRL). The lower portion of the figure shows example methods or areas of future work to model the changes that are possible as the technology approaches commercialization. The top portion of the figure presents examples of technology characteristics that could change throughout the development and deployment process, which can each be assessed using different scale-up models from the literature, and is adapted from van der Hulst, et al. (2020) and Weyand, et al. (2023). These learning effects roughly scale with technology stage, indicated with the arrow. Note that Figures 2–7 are illustrative: they represent feedback from workshops and author discussions, rather than prescriptive guidance or definitive consensus, while curve shapes and heights are qualitative, guided by the experience and expertise of co-authors.*

The key challenge lies in selecting and motivating use of appropriate scale-up methods. General frameworks like those proposed by Tsoy et al. (2020) and Moni et al. (2019) offer cross-sector guidance, but technology-specific methods may be needed and exist for some sectors, for example in heat production and heat pumps (Caduff et al., 2014). Tailored approaches, while less generalizable, yield likely more accurate projections (Piccinno et al., 2016). Some scaling relationships are domain-specific; e.g., biochemical processes require different kinetics than chemical processes (Elginoz et al., 2022). The lack of universal laws should not discourage the use of scale-up methods altogether; rather, it highlights the need for careful contextual application and transparent communication of limitations to avoid misinterpretation.

Applying scale-up frameworks enables LCA practitioners to generate initial point estimates or uncertainty-bounded ranges of future impacts, aiding stakeholder decisions and guiding technology developers when design changes remain feasible. Data-driven advances bolster this capacity: for example, Fozer et al. (2025) demonstrated how experimentally-derived learning rates can scale LCI data for electrochemical ammonia production, replacing arbitrary assumptions. Machine learning could improve scale-up accuracy in process simulations, for example by identifying critical inputs like energy use in novel polymer and chemical manufacturing (Blanco et al., 2024). Overall, incorporating scale-up enhances LCA relevance and supports targeted R&D to improve environmental performance and competitiveness against incumbents.

## Opposition to resolution

Early-stage technologies are often not yet fully is defined, with speculative data on efficiency, materials, and system boundaries, while frequent redesigns limit projection reliability. This makes credible scale-up modeling difficult (Pizzol et al., 2021). For example, LCAs of perovskite solar cells have included scale-up modeling, but it is often applied to materials and processes with high-performance only at the lab-scale, such as solution-synthesized $CH_3NH_3PbI_3$ or Spiro-OMeTAD (Çelik et al., 2016). These systems are among the most studied, yet stability and scalability challenges make them unlikely commercial candidates. Meanwhile, actual high-TRL-proven materials and processes are seldom disclosed due to proprietary constraints. As a result, it is unclear whether the well-studied lab systems are appropriate proxies, and including scale-up modeling may convey false certainty or even introduce additional error. The use of analytical upscaling for emerging technologies with questionable viability and known scale-up barriers therefore risks misrepresentation of future sustainability impacts.

Scale-up processes are rarely linear. Equipment, energy sources, material purity, and integration evolve unpredictably. Initial LCAs of quantum computers underestimated environmental impacts by ignoring error correction overhead, cryogenic cooling, and nonlinear infrastructure growth (Cordier et al., 2025). Common correction factors (e.g., 0.1–0.3 lab-to-industry energy use) risk oversimplifying complex systems (Piccinno et al., 2016; Turconi et al., 2013). Furthermore, misapplied assumptions—such as equating lab electricity with the demands of industrial heat—can distort outcomes. Many learning curve forecasts have proven unreliable in hindsight; for example, IEA's solar projections have consistently underestimated deployment rates due to misjudged process improvements and economies of scale (Lopez et al., 2025).

Unjustified scale-up models risk biasing decisions. Without standards, studies may rely on optimistic (or pessimistic) assumptions, skewing results amid unknown learning effects and regulatory processes. At very early stages, simpler approaches may be preferable such as using empirical data, simulation outputs, expert or vendor input, or literature proxies. Avoiding premature upscaling keeps LCAs manageable and allows refinement as reliable industrial data emerge. Any effort to "scale up" from laboratory data suggests that commercial-scale performance *can* be predicted and encourages comparison to commercial technologies in a way that is often dubious (see also Topic 4). Especially in cases where the goal is technology progress monitoring, hotspot analysis, or comparison among equivalent TRL options, this goal can likely be met without introducing the added complexity and uncertainty of scaling.

### Common ground

No single, consistent scale-up framework exists, but diverse approaches are advancing such as engineering models, empirical scaling, and scenario analyses. More consistent and accessible methods would help practitioners evaluate technologies holistically rather than in isolation. In short, early scenario modeling can help identify impact hotspots and inform better, environmentally focused design and R&D decisions, but when applied assumptions and limitations must be clearly documented.

Like uncertainty assessments, scale-up modeling should be iterative, accounting for TRL, technology type, and context. Distinct challenges arise between stages: TRL 2 to 6 may involve process changes, while TRL 6 to 8 focuses on optimization and integration. Clear guidance is needed on applying general versus technology-specific methods. Better coordination between chemical engineering simulations and LCA scale-up methods would enhance realism. We echo the call from the community for an open-source repository of LCA and process-specific upscaling inventory data (Tsoy et al., 2020; Van Der Hulst et al., 2020), providing examples and proxy trajectories to mitigate data scarcity. Agreement on appropriate methods and user-friendly workflows would accelerate adoption and improve consistency among LCA practitioners.

## Topic 4: Comparison with incumbent

LCA is increasingly used in technology advancement decisions across public and private sectors (Jegen, 2024). These decisions often rely on comparisons between emerging and incumbent technologies. Here, "incumbent" refers to the dominant or currently used technology (also termed “reference,” “benchmark,” or “baseline”) typically the one with the largest market share or the system that the new technology aims to replace. Despite its importance, conducting representative comparative LCAs between emerging and incumbent technologies is challenging due to uncertainties in selecting appropriate incumbents, deployment timing, maturity, and functional equivalence. Some practitioners argue that these uncertainties can render comparisons misleading, advocating instead for hotspot analysis and deferring comparative assessments until more robust data is available (Cucurachi & Blanco, 2022; Van Der Giesen et al., 2020).

Because emerging technologies are future facing, their functions and contexts of use may be uncertain, and today’s incumbents might not reflect future deployment conditions. In such cases, comparisons based on current technologies may understate benefits if they fail to consider potential improvements in the emerging system. At the same time, the benefits might be overstated if compared to current incumbents assuming those do not also improve or evolve over time. Sometimes, no single incumbent or commercial benchmark exists, requiring the construction of a reference mix approximating the intended function. When data are sparse or uncertain, comparisons may be limited to partial assessments using only a select set of performance metrics, such as marginal cost or best-in-class environmental impact.

Differences in technological maturity and production scale also complicate comparisons—for example, between pre-commercial pilots and optimized industrial facilities—introducing asymmetries in assumptions and potential inconsistencies in methodological choices. Moreover, the function or intended application of an emerging technology may evolve over time, necessitating iterative reassessments of the relevant comparator (Moni et al., 2019). Sodium-ion batteries, for instance, were originally targeted for high-energy applications but are now known to be better suited to stationary storage and short-range electric vehicles (Peters et al., 2016). Similarly, $TiO_2$ nanomaterials serve multiple roles, from coatings to electronics, with varying performance metrics and environmental profiles, complicating comparative LCAs (Hischier et al., 2017).

**Figure 5** outlines how confidence in selecting a comparator and the value of comparisons evolve as technologies mature. However, few LCA frameworks offer concrete guidance for comparison modeling. Standards like ISO 14040 lack specificity, leaving ambiguity at early TRLs. In specific domains, such as $CO_2$ capture and utilization, guidelines recommend aligning system boundaries with deployment timelines and incorporating scenario analysis to capture uncertainty and learning (Langhorst et al., 2022; Skone et al., 2022). These methods, while helpful, are usually tied to particular applications or maturity timelines and are not broadly generalizable. Scenario-based projections also remain inherently subjective and context-dependent in their assumptions, and no consensus has yet emerged on best practices. Examples include those discussed in the uncertainty section: Blanco et al. (2020), Douziech et al. (2023) and Grimaldi (2020) all propose different approaches to improve comparisons against incumbents.

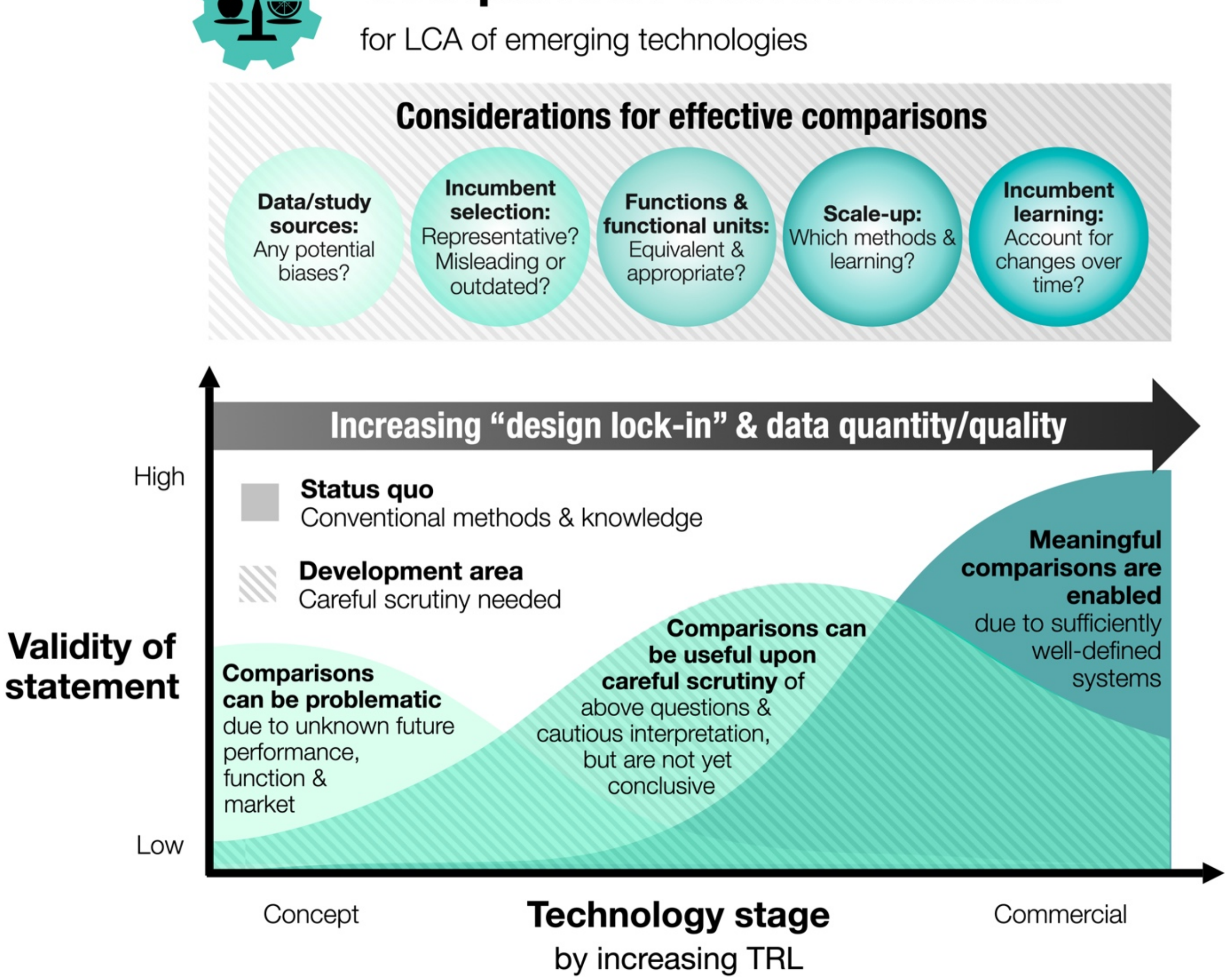

***Figure 5. Challenges and opportunities of making comparisons with incumbent technologies across technology readiness levels (TRLs).*** *The top panel presents key questions practitioners and decision-makers should consider enabling appropriate comparisons, including data sources, incumbent selection, functional equivalency, scaling approaches, and accounting for changes in incumbent technologies over time. The hashed area in the upper panel represents our view that these considerations warrant greater emphasis moving forward. The bottom panel illustrates schematically how the ability to make and interpret comparisons evolves with technology maturity. At very early stages, high uncertainty about the future performance of both emerging and incumbent technologies may make meaningful comparisons inappropriate. As technologies advance, more robust comparisons incorporating uncertainty analysis become possible, so long as comparisons are treated with careful scrutiny and the questions in the top panel are addressed. This stage is hashed to indicate an area where clearer LCA guidelines could support more effective comparisons. Closer to commercialization, systems become sufficiently well-defined to enable meaningful benchmarking with incumbents. At the same time, the potential for both design and technology "lock-in" increases as the degrees of freedom reduce. Note that Figures 2–7 are illustrative: they represent feedback from workshops and author discussions, rather than prescriptive guidance or definitive consensus, while curve shapes and heights are qualitative, guided by the experience and expertise of co-authors.*

Controversial resolution: LCAs of emerging technologies should be compared to incumbent technologies

## Support for resolution

Technologies destined for application in an existing market will have to outperform the incumbent if any environmental benefits are to be achieved, inherently suggesting a need for comparison. LCA results that focus on relative differences between alternatives can be meaningful even when uncertainty is high. For example, identifying hotspots is valuable when conducting an LCA of a system in isolation; however, it does not provide sufficient information to benchmark or measure progress in a future deployment scenario in the way that a comparison does. As noted previously, one argument for implementing LCA early in the development cycle is that changes are more cost-effective at this stage than in mature systems (Cucurachi et al., 2018). However, without effective benchmarking through comparison, it is difficult to evaluate improvements over the course of development.

There is also value in comparing multiple variations of a proposed technology against a singular baseline. For example, developers may be evaluating a variety of potential options within a given technology space, and consistent benchmarking facilitates more useful comparisons within variants (Heijungs et al., 2019). Comparative LCA can also be useful when evaluating multiple potential functions of an emerging technology relative to a variety of different potential end uses and associated incumbent technologies. The results provide decision makers with a comprehensive context of the potential impacts of the emerging technology, which can allow for better assessment of potential trade-offs when choosing between the emerging technology system and existing incumbents.

## Opposition to resolution

At early development stages, data limitations constrain LCA quality. Certainty about end uses of emerging technologies may be incomplete, making it difficult to select an appropriate functional unit. Key parameters—such as co-product handling, recycling feasibility, or end-of-life impacts—may be poorly understood. This makes identifying a truly equivalent comparator extremely difficult.

Even in cases with existing guidance (e.g., Langhorst et al., 2022; Skone et al., 2022), comparator selection remains nuanced. LCA studies often lack a clear definition or justification for the chosen reference system. Depending on whether a marginal or average system is used, or how optimistic future performance assumptions are, the same emerging technology can appear environmentally beneficial or detrimental. This inconsistency can confuse stakeholders and risks misleading investment or policy decisions. It also opens the door to bias, where LCA practitioners may (intentionally or not) "cherry-pick" data or comparators to favor a desired outcome.

Outdated or static background data compounds the problem (Cucurachi & Blanco, 2022). Consider solar PV: many LCAs still benchmark against multi-crystalline silicon using data from as far back as 2005 (Cucurachi & Blanco, 2022; Müller et al., 2021). Even the most recent PV LCI guidelines (Danelli et al., 2024; Smith et al., 2024) focus on technologies that, while dominant until 2023, are expected to account for less than 20% of the market by 2025 and may become obsolete by 2029 (Fischer et al., 2025). Thus, even in sectors with detailed LCI guidance, inappropriate incumbent selection can lead to misleading conclusions, likely underestimating the impacts of the emerging technology. Therefore, it may be more responsible to defer comparison until there is sufficient data to support robust conclusions.

## Common ground

When conducted carefully and transparently, early-stage comparative LCAs can reveal environmental trade-offs, inform performance targets, and support R&D prioritization. However, comparisons at low TRLs must be interpreted cautiously (see considerations in Figure 5), since they rely on uncertain assumptions and preliminary data. Results are often sensitive to choices around system boundaries, background datasets, and functional units. Incumbents generally represent systems refined through years of optimization, while emerging technologies may only exist at bench or pilot scale.

The ultimate goal of comparative LCA is to inform decisions—whether for R&D direction, investment allocation, or market strategy. These decisions must not hinge solely on comparative results unless uncertainty has been adequately assessed and communicated. LCA practitioners should work to present degrees of certainty clearly (see Table 1), taking extra care to define the goal of the comparison, consider scale-up effects, and ensure functional equivalence. Comparator data should be updated, and care must be taken to reflect potential future deployment contexts. Emphasis should be placed on rigorous, transparent quantitative reporting rather than normative or interpretive conclusions. Simultaneously, decision-makers must scrutinize evidence, question assumptions, and align interpretations with their own risk tolerance and strategic priorities. Comparator selection should reflect not only current systems but also evolving market shares, technology lifespans, and infrastructure trajectories.

Final “go/no-go” decisions should be contingent on sufficient data quality and robustness of analysis. The threshold for making such decisions is subjective and rests with the decision-maker, not the practitioner. To support effective comparisons, we advocate for development of guidelines tailored to emerging technologies. This would offer clarity on when comparisons are appropriate, how incumbents should be defined, and how to account for uncertainties in projecting both future performance and evolving benchmarks.

# Topic 5: Standardization

The ISO 14000 series of standards, including ISO 14040 (Principles and Framework) and 14044 (Requirements and Guidelines), provides foundational guidance on conducting environmental life cycle assessment (LCA). Additional requirements stem from various sources, including regulations (e.g., California Air Resources Board, 2020; Cheng et al., 2023), industry consortia (Langhorst et al., 2022), product category rules (e.g., Institut Bauen und Umwelt e.V., 2023), funding agencies (Skone et al., 2022), and frameworks like the European Commission’s Product Environmental Footprint (PEF) (European Commission, 2021, 2025). These standards aim to ensure consistency and quality. When an LCA is labeled “ISO-compliant” or “standardized,” it conveys a veneer of rigor and comparability. However, even an ISO-compliant LCA allows broad latitude in how an LCA is conducted including modeling assumptions, data, and uncertainty assessment techniques employed.

Consider, for example, the following guidance on data collection for LCA (excerpted from ISO 14044): “Data selected for an LCA depend on the goal and scope... In practice, all data may include a mixture of measured, calculated or estimated data.” (ISO, 2006b) Practitioners are given broad discretion in selecting appropriate data while acting in good faith to follow the standard. The result is that practitioner choices can have an enormous impact on results. For example, measuring flows onsite versus estimating them using proxy or calculated data will likely lead to different system boundaries and results. ISO 14040 reinforces this flexibility: “There is no single method for conducting LCA. Organizations have the flexibility to implement LCA ... in accordance with the intended application and the requirements of the organization.” (ISO, 2006a)

While flexibility is an important feature of application-agnostic standards, it can also create a false impression of harmonization and comparability across LCA studies, particularly at low TRLs, where uncertainty is greatest. In these cases, claims of “ISO compliance” may obscure significant modeling assumptions and sources of uncertainty that would not be relevant to a mature technology assessment (see **Figure 6**; Bergerson et al., 2020). Although ISO calls for uncertainty and sensitivity analyses for public LCA comparisons, it lacks specific guidance for low-TRL applications, making it difficult to balance methodological rigor with perceived comparability and robustness.

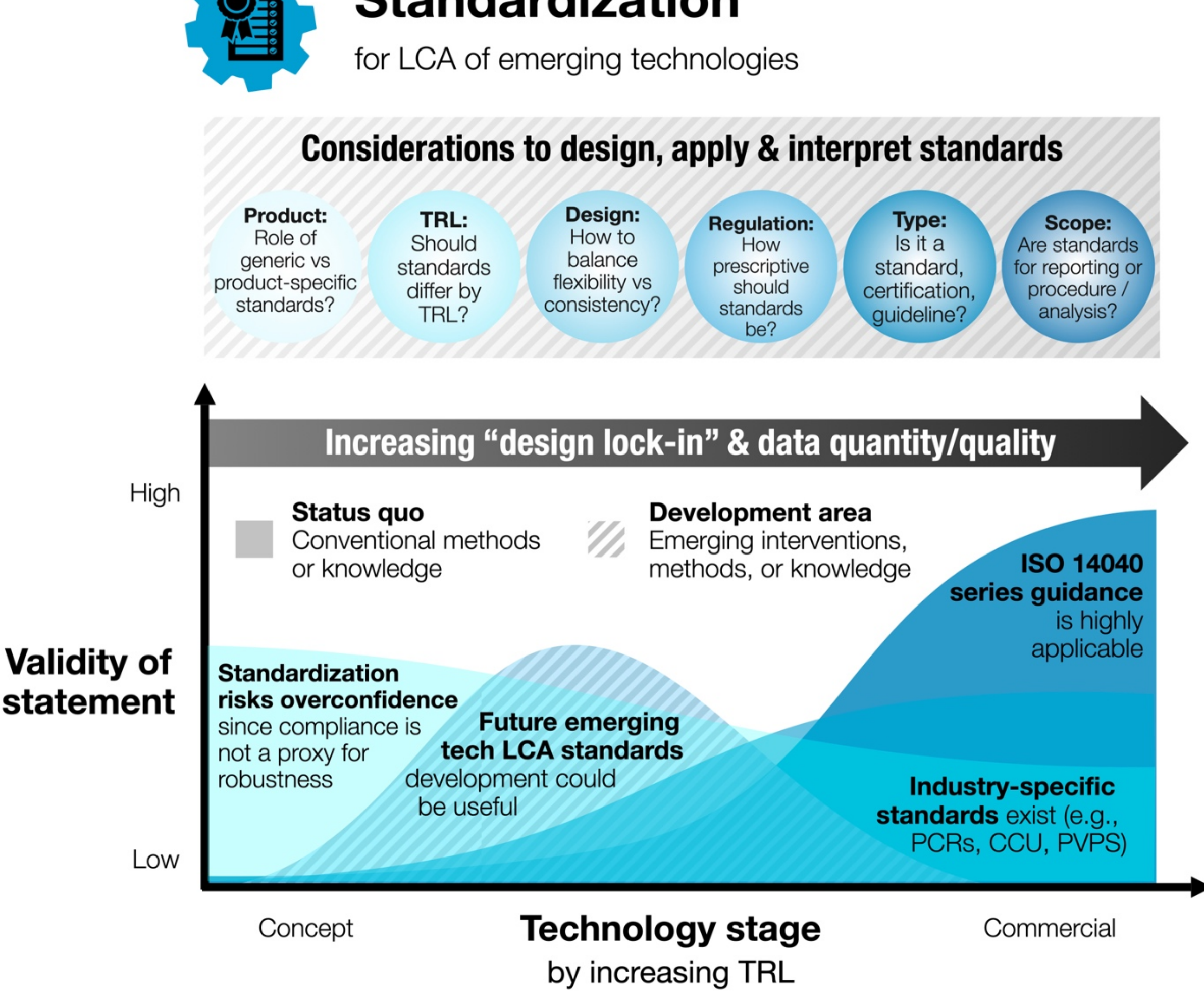

***Figure 6. Standardization options for LCA of emerging technologies.*** *Across technology development stages, as well as across industries, specific characteristics and methods that are and can be standardized differ. Similarly, standardization brings about unique advantages and disadvantages at various stages and fields. The top of this figure highlights key questions LCA practitioners and stakeholders should be considering when developing, using, and interpreting standards. The bottom of the figure represents qualitative considerations as a function of TRL, and existing and proposed standards and guidelines are also highlighted. We highlight several example standards: PCRs - Product Category Rules (British Standards Institution, 2019)*

*(available for many technology product categories); CCU – Carbon Capture and Utilization (Langhorst et al., 2022); and PVPS - Photovoltaic Power System Program (International Energy Agency, 2017). It is important to note that these are representative approaches and existing examples and are not intended to be comprehensive. The hashed shading represents the specific opportunity highlighted in this paper to expand and improve standards during the mid-development period. Note that Figures 2–7 are illustrative: they represent feedback from workshops and author discussions, rather than prescriptive guidance or definitive consensus, while curve shapes and heights are qualitative, guided by the experience and expertise of co-authors.*

Early-stage LCA guidance exists, but it is fragmented. For instance, the EU JRC ILCD Handbook promotes default scenarios and pedigree matrices (European Commission, 2010), and the UNEP/SETAC Social LCA guidelines encourage iterative assessments and stakeholder engagement (UNEP, 2020). The PEF framework allows for use of proxy data and modular datasets and encourages data quality scoring and gap-filling techniques but was designed for commercial products and is less applicable to emerging technologies.

More detailed guidance exists in specific industries (lower right curve in Figure 6). The carbon capture sector, for example, includes the EPA's DACS guidelines and the Global $CO_2$ Initiative's CCU framework (Langhorst et al., 2022). The U.S. Department of Energy's (DOE's) Biomass Carbon Removal and Storage (BiCRS) best practices recommend cradle-to-grave system boundaries, biogenic carbon accounting, and scale-appropriate modeling (Cai et al., 2025). For nanomaterials, the OECD and NanoReg2 suggest tiered, qualitative approaches and screening tools (OECD, 2015; Zondervan-van den Beuken, 2014). The IEA's PV Task 12 offers prospective data use and "lab-to-fab" substitutions (International Energy Agency, 2017; Müller & Wambach, 2017). However, coverage and depth are inconsistent across sectors.

Moreover, generalized standards for early-stage technologies are lacking but could improve consistency and method clarity (second curve from the left in Figure 6). While no standard can cover all contexts or substitute for practitioner judgment, standards tailored to specific domains or funding calls could improve reproducibility. Still, there are risks: standard development processes can be politicized or biased (Blind & von Laer, 2022), and overly rigid standards may oversimplify or misalign with evolving applications.

### Controversial resolution: Specific standards are needed for LCA of emerging technologies.

## Support for resolution

Standards provide explicit guidance for methodological steps, definitions, and assumptions that might otherwise be left to the judgment of individual practitioners. However, in many cases—such as estimating scale-up parameters—informal norms substitute for documented or detailed best practices, creating space for inconsistency. In the absence of empirical data, early-stage LCAs often borrow values from incumbents, yet guidance on when this is appropriate is limited. For example, LCAs of new battery chemistries commonly use lithium-ion lifetimes as proxies, likely biasing results due to high LCIA sensitivity to cycling (Peters et al., 2017; Peters & Weil, 2018). Standards could guide estimates in pre-commercial systems where flows cannot be directly measured. They could also help discourage flawed practices such as misusing scaling factors to infer industrial-scale estimates from lab data (e.g., energy use), a widespread practice but one highly sensitive to assumptions (Majeau-Bettez, 2021; Piccinno et al., 2018; Röder et al., 2022). The DOE's DACS guidelines caution against such arbitrary scaling, but this warning may go unnoticed outside the field.

Standards could mitigate scenarios where disagreements on best practices are common, especially at low TRL. LCAs of emerging solar cells, for instance, often yield inconsistent results due to varying system boundaries, functional units, assumed scale, and end-of-life treatment (Cellura et al., 2024). In other cases, novel technologies challenge the use of existing methods, prompting entirely new approaches. For example, Biomass Carbon Removal and Storage (BiCRS) technology relies on the photosynthetic accumulation, transformation, and storage of biomass – processes that when combined together have a lack of historical analogs, spurring new carbon accounting protocols (Cai et al., 2025). The result degrades the insights derived from the analysis as methods applied vary so widely that many studies stand alone as independent and noncomparable, and we see this as the unfortunate reality for many LCAs of emerging technologies.

For these reasons, we argue that creating emerging technology standards is necessary not only to ensure methodological coherence across tiers, but also to support reproducibility and comparability across studies. Such efforts could also facilitate LCA education and training for novice analysts, researchers, product engineering and design teams, funders, and other stakeholders involved in early-stage technology development, and provide transparency for decision-makers interpreting results.

## Opposition to resolution

Overly prescriptive standards may do more harm than good. It is difficult to create harmonized standards that are sufficiently detailed to be useful but general enough to accommodate different technologies. Consensus-building is also challenging, particularly when stakeholders have conflicting technical or policy goals (Huijbregts 2014). Rigid standards may also restrict the expert judgment required to make thoughtful, case-specific modeling choices, especially in emerging technologies where novelty and data scarcity are the norm. Even sound standards may create false confidence in LCA findings (see **Figure 5**). "Compliance" can mask modeling subjectivity or bias, and many decision-makers may not fully appreciate what a compliant label entails.

Furthermore, standardization does not guarantee accuracy. LCA choices—like system boundary setting, selection of functional unit and co-product allocation method—require contextual, often iterative judgment. The more data-poor, the more subjective these choices become. For example, in systems with uncertain or evolving supply chains, assumptions about input sourcing or end-of-life treatment often reflect informed guesses rather than empirical data. While using a single set of rules (e.g., energy-based allocation) can support consistency, it may not be the most appropriate choice especially for low TRL processes; exploring alternatives (e.g., system expansion or economic-based allocation) and sensitivity analyses could yield more valid results. Over-standardization risks discouraging LCA adoption by introducing unnecessary complexity or cementing inadequate practices. Furthermore, standards can potentially entrench context-dependent value choices as requirements. Since value systems and norms are often not consistent across technology areas, geographies, and political environments, prescriptive standardization risks the improper prioritization of comparability over flexibility (Bach et al. 2018).

## Common ground

While there is disagreement about how and what to standardize, there is consensus that improving transparency through standardized reporting would be beneficial. Table 1 outlines recommended elements for such reporting. Important items of emphasis include the TRL of the technology and whether the inventory represents the technology "as is" (at its current state of development) or if some modification of data has been conducted to model the technology "as might be" at commercialization (and whether these modifications represent central, optimistic or pessimistic cases). Clarity is also needed regarding assumptions around the future conditions (e.g., background systems) into which the technology may be deployed, which may include internally consistent standard scenarios such as those proposed by Sacchi et al. (2022).

At a minimum, we recommend a critical review of harmonized guidelines for LCA of emerging technologies across industries and regulatory bodies. Where full standardization isn't feasible, transparency about methods (including which aspects align with standards, and which are unique assumptions or customized methodological extensions) becomes vital. We direct LCA practitioners to the considerations shown in the upper panel of Figure 6, and we suggest focused guidance development around key questions such as:

1. **Rigor criteria**: What separates life cycle thinking from a comprehensive and robust LCA?
2. **Methods**: Which are most appropriate for different TRLs (e.g., for scale-up)?
3. **Foreground data**: When should measurement be required versus projection?
4. **Data quality**: How should key dimensions—temporal, geographical, and technological representativeness; completeness and reliability—be reported?
5. **Uncertainty**: How should modeling uncertainties and biases be communicated?
6. **Compliance**: What should "compliance" with LCA standards mean for emerging technologies?

Ultimately, standardization must allow flexibility. "Shall" statements (for consistency) should be complemented with "should" or "may" statements for sensitivity testing and extensions (e.g., as found in Langhorst et al., 2022). These efforts can be enhanced through broader initiatives such as standardizing background LCI data for future scenarios (e.g., U.S. Department of Energy standard scenarios for the U.S. electricity sector (Gagnon et al., 2024)), which would strengthen consequential modeling across applications, including emerging technologies.

## Topic 6: Stakeholder engagement

LCA of emerging technologies is often conducted to inform high-stakes, urgent decisions (Ravetz & Funtowicz, 2015). However, engagement with community stakeholders is rarely incorporated directly, leading to assumptions about environmental and social priorities that may not reflect community concerns. Ideally, high-stakes LCAs would involve an "extended peer community" with transparent opportunities for public participation. Stakeholders may include subject matter experts, policymakers, investors, industry partners, users, impacted community members, and the broader public. Engagement can vary from informing and consulting to involving, collaborating, and ultimately empowering (Kujala et al., 2022).

Although the ISO 14040 series guidelines do not explicitly mention stakeholder engagement, it has been proposed as a component of "anticipatory LCA" to support responsible research and innovation (Wender et al., 2014). The ISO guidelines require expert review for comparative public assertions (ISO, 2006a, 2006b), which presents one possible channel for engagement. In energy systems modeling, broader frameworks exist to incorporate diverse perspectives on possible and preferred futures (McGookin et al., 2024). **Figure 7** illustrates where different stakeholders can contribute to LCA and environmentally informed design across the development timeline calling attention to specific areas of involvement and areas for further growth. However, meaningful engagement is not without challenges, and its benefits must be weighed against practical limitations, potential biases, and possible risks of involvement.

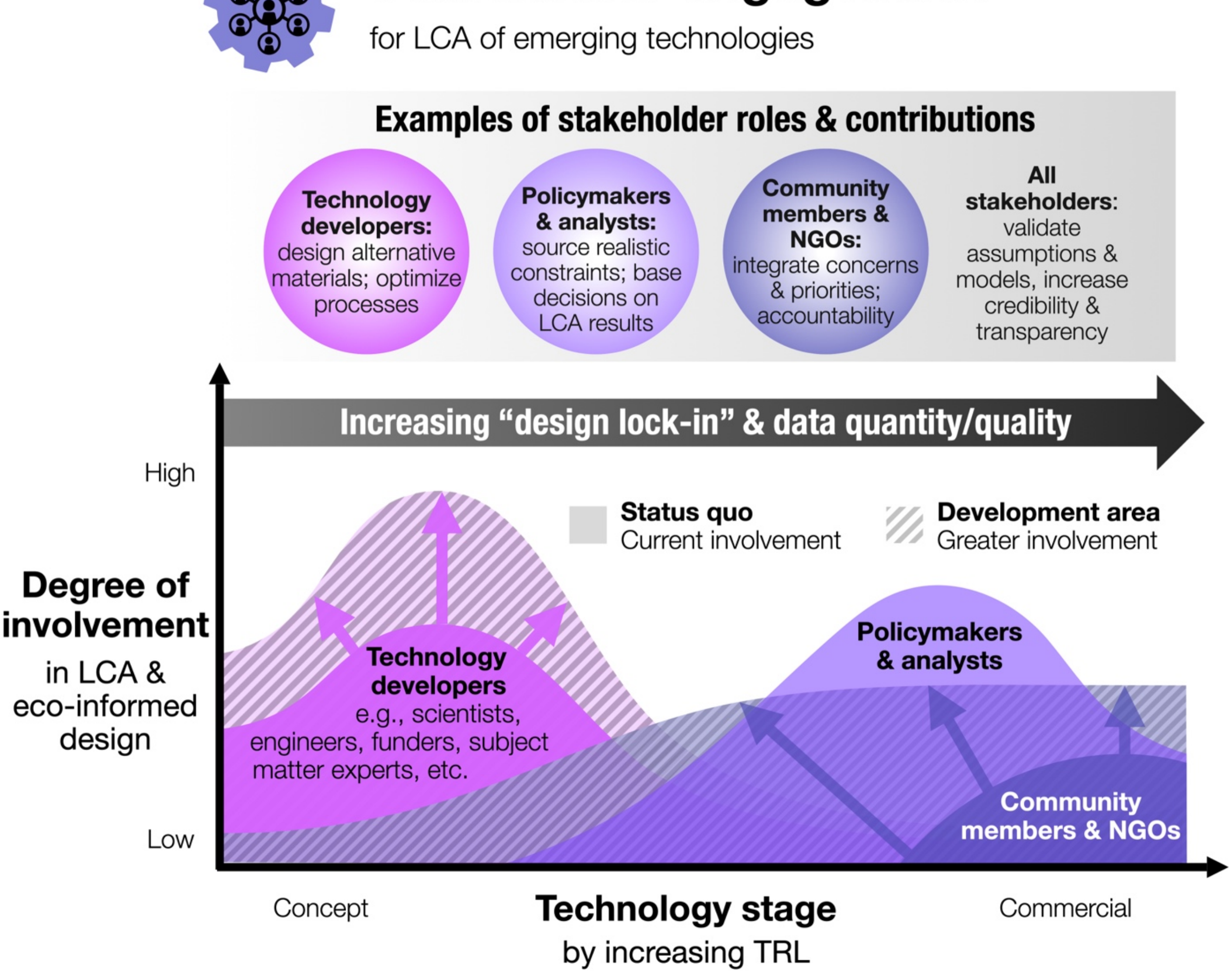

***Figure 7. Stakeholder Engagement in LCA.*** *Key groups of stakeholders at different stages of development can play unique roles in interfacing with the LCA process, with examples of such roles depicted in the upper panel. The involvement roles across three key groups are represented schematically in the bottom panel with curves encompassing various technology readiness levels (TRLs), and ranging from scientists, engineers, and funders at early stages to analysts and policymakers at mid-stages, and community members and environmental NGOs at later stages. The hashed shading represents areas we have identified where engagement can particularly be improved and expanded in the future. We highlight the role of technology developers especially at the earliest stages as a large area of growth, as indicated by the prominent peak. Our consensus is that community members can be engaged throughout the technology development process, which can iteratively inform LCAs throughout. Effective stakeholder engagement—rooted in collaboration and empowerment—can positively impact early stage LCA in myriad ways such as informing relevant topics, improving quality assurance of the results, and facilitating better dialogue across the range of stakeholders potentially interested in an emerging technology. Note that Figures 2–7 are illustrative: they represent feedback from workshops and author discussions, rather than prescriptive guidance or definitive consensus, while curve shapes and heights are qualitative, guided by the experience and expertise of co-authors.*

Controversial resolution: LCAs of emerging technologies should involve diverse stakeholders early and throughout.

## Support for resolution

Stakeholders can offer perspectives that inform not just the design, but the entire development cycle of an LCA. Figure 7 highlights two key groups: subject matter experts and members of society including workers, local communities, and actors along the value chain (UNEP, 2020). These individuals can offer data unavailable in conventional inventories and contextualize assumptions and scenarios. Especially for emerging technologies, which often lack established user bases or community experience, stakeholder engagement during the scoping phase can help define system boundaries, impact categories, and functional units, or prioritize the study's focus to avoid unnecessary costs. Engagement can also inform what performance targets should be achieved before design lock-in occurs.

The inclusion of community stakeholders in LCA at early stages of development can have direct implications for environmental justice and social impact (Cowell et al., 2002; Mathe, 2014). Their inclusion can reduce the dominance of perspectives from developers or analysts. It can also ensure the assessment is responsive to local needs. For example, community members might prioritize air quality over climate change impacts when evaluating a fuel switch from natural gas to biomass (Pitre et al., 2024). Incorporating this insight can broaden the study's relevance and credibility. Moreover, input from stakeholders enhances confidence in LCA models and informs assumptions used in future scenario-building (Sleep et al., 2021).

Subject matter experts possess detailed knowledge of scientific mechanisms, regulatory environments, infrastructure, supply chains, and public sentiment. Their insights are critical to defining feasible use cases, scaling parameters, and selecting proxies, all of which shape the LCA's accuracy. Experts can also offer forward-looking context often missing from the literature (Verdolini et al., 2018). Their critical review functions similarly to formal peer review by highlighting when default or easily accessible data choices may not be appropriate. For instance, recent engagement efforts provided evidence for the need for geographically localized emissions factors rather than regional averages (Sleep et al., 2021). In emerging technology areas, a single LCA study may become a benchmark used by many and could influence policy, funding and technology development decisions, so engaging stakeholders from the outset is important.

## Opposition to resolution

Despite its benefits, stakeholder engagement presents real challenges. ISO standards do not require it (other than critical review), and formal methods for conducting it are underdeveloped. LCA practitioners may lack the training or resources to facilitate meaningful participation, leading to potential unintended stakeholder concerns, annoyance, and delays in completing the LCA projects (Cowell et al., 2002). Identifying appropriate stakeholders is especially difficult for early-stage technologies with undefined or evolving use cases. This opens the door to selection bias and inconsistent outcomes.

Determining which stakeholders to include and what role they should play is a key challenge. Comprehensive consultation is infeasible, and selectively choosing participants may privilege dominant or well-resourced voices while omitting others. Divergent values and priorities across stakeholder groups may further complicate engagement, requiring co-creation rather than consultation alone. Relationship-building takes time and resources. As such, engagement may inadvertently impose burdens—particularly for marginalized communities—which raises ethical concerns. Without care, engagement risks becoming superficial, tokenistic, or even exploitative (Gorman, 2024; Poiner & Drake, 2021; Reed, 2008).

Identifying qualified subject matter experts is equally challenging. In some fields, the expert base may be shallow, and self-proclaimed authorities may lack necessary qualifications. There's also a tendency to consult those with resources and institutional access, often favoring incumbents and sidelining underrepresented perspectives (Suter & Cormier, 2016). Even when experts are available, confidentiality concerns may limit their willingness to share proprietary data or future projections, reducing the utility of their input.

Moreover, not all funders of early-stage LCA may be convinced of the importance of stakeholder engagement, especially if it delays delivery of study results beyond the decision-making window. Concerns over intellectual property, reputational risk, or the possibility of critical public feedback can make funders or developers hesitant to participate in open engagement. In some cases—such as exploratory hotspot analyses, internal assessments of greenhouse gas emissions, or process changes with minimal impact on end-users—stakeholder involvement may not be necessary. Transparent communication of assumptions, limitations, and uncertainties might be sufficient in these cases.

### Common ground

Despite differing views on the timing and intensity of stakeholder engagement, there is broad agreement in our group that stakeholders can contribute insights into community priorities, geographic context, market dynamics, consumer behavior, and material and energy constraints. Subject matter experts can provide grounded knowledge that can strengthen the validity of assumptions and study design. However, stakeholder engagement must be intentional, well-resourced, and tailored to the LCA's goals and stage of development. Table 1 and Figure 7 summarize our shared takeaways.

While Figure 7 depicts general stakeholder groups and qualitative technology stages, consensus was not reached on prioritization or sequencing of specific stakeholder actions, but we do agree that the engagement process should always align with the study's objectives. Even when comprehensive and robust LCAs are not pursued, stakeholders (including community members and developers) can co-create alternative scenarios to explore sustainability implications. However, guidance is lacking on how to approach LCA stakeholder engagement depending on study type, TRL stage, stakeholder group, or impact category. Developing clear frameworks to guide this engagement would serve the LCA community well to produce more accurate LCAs that better inform decision-makers.

Looking ahead, Figure 7 identifies opportunities to expand and deepen engagement. Early involvement of technical experts at low TRLs and consistent participation of affected communities across all TRLs can improve transparency, relevance, and equity. Case studies of LCA with stakeholder input could provide templates for effective practices and clarify appropriate levels of engagement, from informing to empowering. Promoting community dialogue and clear role definition can build trust and help avoid tokenism. Additionally, LCA practitioners must remain aware of their own biases and commit to early, inclusive, and purposeful engagement to support more robust and just assessments of emerging technologies.

## Conclusion

This study has critically examined controversial topics associated with the application of LCA to emerging technologies, using a Faraday-style approach to consider dissenting opinions among our expert group, areas of consensus, and next steps for the field. We have reviewed six key methodological debates and open questions in LCA (see Figure 1):

1. **Appropriate use of LCA**: How can LCA or life cycle thinking methods be selected and communicated to reflect the unique characteristics of emerging technologies to ensure appropriate interpretation? (Figure 2)
2. **Uncertainty assessment**: How can and should uncertainty be quantified and managed in the context of limited data? (Figure 3)
3. **Scale-up**: What are the most effective ways (if any) to assess potential future environmental and social impacts in hypothetical commercial settings? (Figure 4)
4. **Comparison with incumbents**: What are the best practices for fairly conducting (or avoiding) comparative LCAs of emerging technologies? (Figure 5)
5. **Standardization**: What gaps exist in current standards, and how might they be revised or augmented to accommodate early-stage technologies? (Figure 6)
6. **Stakeholder engagement**: How can LCA practitioners, technology experts, and society collaborate more effectively? (Figure 7)

These questions can serve as a launching point to explore the complexities of LCA for new and evolving technologies, with several thematic synergies emerge (see Table 1, summarizing outcomes, recommendations, and next steps from our network's discussions). Within the LCA community, particularly in the context of emerging technologies, there is sustained attention to methodological development. Core concerns include addressing scale-up challenges, managing uncertainty, and refining comparative analysis. The community is also working to foster effective collaboration among practitioners, technology developers, and impacted stakeholders to improve LCA practices.

Another theme across all six topics (see Table 1) is the need for clear communication of data and study quality, covering both foreground and background data, compatibility with data quality matrices, temporal and geographical resolution, and links to uncertainty characterization. Across our network, there is a shared view that the interpretation and communication of LCA results must be handled carefully, especially given the high uncertainty inherent to emerging technologies. Results should be considered provisional insights that help guide research and development, rather than definitive predictions. However, further clarity is needed regarding how an LCA's robustness should be defined, conveyed by practitioners, and interpreted by stakeholders.

Our discussions revealed both consensus and disagreement across the six key themes. Agreed-upon points include the importance of life cycle thinking for decision-making, the need for transparent communication of limitations, and the value of understanding technological nuances and uncertainties. We agree on the need for coordinated efforts to build shared data infrastructure and to critically review existing guidelines to inform future best-practices or standards. However, disagreements persist around when and how to apply LCA during development, appropriate methods for early-stage uncertainty characterization, how to scale up and compare technologies, and whether specific standards should be adopted. We identify key trade-offs—flexibility vs. comparability, early engagement vs. stakeholder burden, transparency vs. comprehensiveness, sophistication vs. overconfidence, and comparability vs. contextual relevance—which underscore the need to balance credibility with practicality in LCA for emerging technologies.

Applied early, LCAs can identify environmental hotspots, prioritize areas for improvement, and flag topics needing further investigation. Practitioners should avoid overstating findings and instead communicate assumptions, uncertainties, and limitations clearly. Scenario-based presentation, outlining ranges of possible outcomes, is essential to convey how varying assumptions affect results. Interpretation should remain consistent with the decision the LCA is intended to inform. Poorly communicated or misinterpreted results risk misguided decisions and public misconceptions; by contrast, clear and contextualized communication can support sustainability-oriented design, funding, and policy.

While debates about these six topics will continue, this paper offers a guide for stakeholders navigating the intricacies and uncertainties of LCA for emerging technologies. We aim to support the LCA community in confronting methodological challenges and foster constructive dialogue around unresolved issues. We also seek to enhance LCA literacy among non-practitioners, empowering them to critically interpret results and participate meaningfully in the assessment process. Our goal is not to prescribe rigid solutions, but to encourage iterative refinement, transparent communication, and inclusive engagement. As technologies evolve, so must our LCA approaches. By continuing to address the uncertainties and trade-offs discussed here, LCA practitioners can ensure their work remains a vital tool in shaping sustainability-informed technology decisions.

## Acknowledgments

We wish to acknowledge the support of the American Center for Life Cycle Assessment (ACLCA) and International Symposium on Sustainable Systems and Technology (ISSST) for hosting workshops held by our *LCA for Emerging Technologies* research network as part of their annual conferences. The authors also appreciate the engagement of the many conference workshop participants who generously contributed their ideas and perspectives.

## Funding Information

**Carbajales-Dale and Chen:** This work was supported as part of the AIM for Composites, an Energy Frontier Research Center funded by the U.S. Department of Energy, Office of Science, Basic Energy Sciences at the Clemson University under award #DE-SC0023389. **Cheng:** This work was supported in part by National Science Foundation Graduate Research Fellowship under grant number DGE2140739. **Moni:** This work was funded by the Department of Energy, National Energy Technology Laboratory an agency of the United States Government, through a support contract. Neither the United States Government nor any agency thereof, nor any of their employees, nor the support contractor, nor any of their employees, makes any warranty, express or implied, or assumes any legal liability or responsibility for the accuracy, completeness, or usefulness of any information, apparatus, product, or process disclosed, or represents that its use would not infringe privately owned rights. Reference herein to any specific commercial product, process, or service by trade name, trademark, manufacturer, or otherwise does not necessarily constitute or imply its endorsement, recommendation, or favoring by the United States Government or any agency thereof. The views and opinions of authors expressed herein do not necessarily state or reflect those of the United States Government or any agency thereof. **Posen:** This research was undertaken, in part, thanks to funding from the Canada Research Chairs Program (CRC-2020-00082). **Wachs:** This work was authored in part by the National Renewable Energy Laboratory, operated by Alliance for Sustainable Energy, LLC, for the U.S. Department of Energy (DOE) under Contract No. DE-AC36-08GO28308. Funding provided by U.S. Department of Energy Office of Energy Efficiency and Renewable Energy Industrial Efficiency and Decarbonization Office. The views expressed in the article do not necessarily represent the views of the DOE or the U.S. Government. The U.S. Government retains and the publisher, by accepting the article for publication, acknowledges that the U.S. Government retains a nonexclusive, paid-up, irrevocable, worldwide license to publish or reproduce the published form of this work, or allow others to do so, for U.S. Government purposes. **Woods-Robinson**: This work was supported by a Distinguished Postdoctoral Fellowship at the University of Washington's Clean Energy Institute. **Abeynayaka**: This work was partially supported by the MINAGRIS project (grant agreement No. 101000407) funded under the European Union's Horizon 2020 Research and Innovation Programme.

## CRediT Statement

**Rachel Woods-Robinson:** Investigation, Writing, Editing, Revising, Visualization (lead). **Mik Carbajales-Dale:** Conceptualization, Investigation, Writing, Editing, Project Administration (co-lead). **Anthony Cheng:** Investigation, Editing, Revising, Visualization. **Gregory Cooney:** Conceptualization, Investigation, Writing, Editing. **Abby Kirchofer:** Conceptualization, Investigation, Writing, Editing, Revising. **Heather Liddell:** Conceptualization, Investigation, Writing, Editing, Revising, Visualization. **Lisa Peterson:** Investigation, Writing, Editing, Revising, Visualization. **I. Daniel Posen:** Conceptualization, Investigation,

Writing, Editing, Revising. **Sheikh Moni:** Conceptualization, Investigation, Writing, Editing. **Sylvia Sleep:** Conceptualization, Investigation, Writing, Editing, Revising. **Liz Wachs:** Investigation, Writing, Editing, Revising, Visualization. **Shiva Zargar:** Investigation, Writing, Editing, Revising, Visualization. **Hao Chen:** Editing and Revising. **Amila Abeynayaka:** Writing, Editing, Revising. **Manish Kumar:** Writing, Editing, and Revising. **Joule Bergerson:** Conceptualization, Investigation, Writing, Editing, Revising, Visualization, Project Administration (co-lead), Corresponding Author.

## Data Availability Statement

Data sharing is not applicable to this article as no new data were created or analyzed in this study.